\documentclass{aa}
\usepackage{graphicx}
\usepackage{supertabular}
\usepackage{amssymb}
\usepackage{amsmath}
\usepackage{amsfonts}
\usepackage{natbib}
\usepackage{txfonts}
\bibpunct{(}{)}{;}{a}{}{,}

\def\deg{$^\circ$}

\begin{document}
\title{Edge-on T Tauri stars\thanks{Based on
observations obtained with {\sc Uves} at the ESO Very Large
Telescope, Paranal, Chile (proposal No.~70.C-0041(A)).}}

\author{Immo Appenzeller\inst{1}
      \and Claude Bertout\inst{2}
      \and Otmar Stahl\inst{1}}

\offprints{I. Appenzeller} \institute{Landessternwarte,
K\"onigstuhl, D 69117 Heidelberg, Germany
    \and Institut d'Astrophysique, 98bis, bl. Arago, F 75014
    Paris, France}

\date{Received March 30, 2004 / Accepted January 8, 2005}

\abstract{Using the {\sc Uves} echelle spectrograph at the ESO VLT
we obtained two-dimensional high-resolution (R = 50 000) spectra of
the edge-on disk objects HH30$^{*}$, HK~Tau~B, and HV~Tau~C. For
comparison purposes we also observed with the same equipment both
the classical T~Tauri star HL~Tau and the  active late-type star LDN
1551-9.  The spectra of all three observed edge-on disks consist of
a T~Tauri emission and absorption line spectrum with superimposed
jet emission lines. Analysis of the spectra confirmed that the disks
are completely opaque at visible wavelengths and that light from the
central objects reaches us only via scattering layers above and
below the disk planes. The central objects of our targets were found
to be normal T~Tauri stars showing moderate but different amounts of
veiling of their photospheric spectra, indicating different
accretion rates or evolutionary stages. We suggest that all
classical T~Tauri stars (CTTSs) show this observed morphology when
viewed edge-on. Part of the jet emission from edge-on systems is
directly visible to us in the forbidden lines as well as in
H$\alpha$ and He\,{\sc i}, a finding which contradicts the present
paradigm of a pure magnetospheric accretion origin for the formation
of hydrogen and helium emission lines in moderately active CTTSs.
From a comparison with those Taurus-Auriga CTTSs for which the
inclination is reliably known, we conclude that the view angle of
CTTS systems is one of the key parameters governing apparent
H$\alpha$ emission strength in the T~Tauri class. We discuss the
various possible formation regions for the Na\,{\sc i}\,D\ lines and
show that profiles similar to observed ones can be formed at the
base of the disk wind. }
\authorrunning{I. Appenzeller et al.}
\maketitle

\section{Introduction}\label{introduction}

The development of high angular resolution imaging techniques in the
optical, infrared, and millimetric bands has led to the discovery over
the last decade of a number of resolved disks surrounding pre-main
sequence stars \citep[see the review by][]{2002EAS.....3..183M}.
These observations have confirmed the long held belief, based on
indirect evidence, that young, solar type stars are often associated
with disks \citep[e.g.][]{1984A&A...141..108A,1988ApJ...330..350B}.

While the circumstellar disks are known to contain the dust and gas
needed for planetary systems to develop, their structure and
evolution remain elusive.  The relationship between these disks and
the ubiquitously observed proto-stellar jets is a related question
that also remains unsolved. A disk/jet connection is well
established \citep[e.g.][]{1990ApJ...354..687C}, but the origin of
the jet and its driving mechanism remain topics of intense debate
\citep[see the review by][]{2002EAS.....3..147C}.

Current models of gravitational collapse \citep{1999ApJ...525..330Y}
predict that a quasi-hydrostatic accretion disk forms around the
stellar embryo and grows in size and mass as the collapse proceeds,
while feeding mass to the central star. However, the computed disk
mass and size as the star approaches the main sequence both appear
larger than observed, possibly because the role of the proto-stellar
jet, of MHD instabilities, and of non-axissymmetric effects in
carrying away angular momentum of the infalling matter is not taken
into account in these simulations.

That the disk should grow radially because of outward angular
momentum transport as its evolution proceeds appears nevertheless
plausible, as already noted by \cite{1974MNRAS.168..603L} and more
recently by \cite{1998ApJ...495..385H}. It is perhaps not
surprising, therefore, that the largest circumstellar disks are
observed around young solar type stars that have become visible in
the near-infrared and optical ranges, the T~Tauri stars or Class II
objects in the evolutionary scheme for solar-mass pre-main sequence
objects originally devised by \cite{1987ApJ...312..788A} and
extended by \cite{1993ApJ...406..122A}.

Class II stars are surrounded by actively accreting disks (Bertout
et al. 1988). The accretion onto the star is thought to be
proceeding along the magnetic field lines of the stellar
magnetosphere, which disrupts the inner disk regions. Class II
objects also harbor jets, although less powerful ones than in
previous evolutionary phases, such that they are often referred to
as micro-jets. Because Class II objects are visible in the optical
range, we can use the powerful technique of high-resolution optical
spectroscopy to study them.

This technique has already been fruitfully used decades ago for the
brightest young stellar objects \citep[e.g.][]{1984ApJ...280..749M}
and, more recently, for several Orion proplyds
\citep[e.g.][]{1999AJ....118.2350H}, the spectra of which are
strongly affected by disk photo-evaporation caused by the intense UV
radiation field from the Trapezium stars. But no high-resolution
spectroscopy of the recently imaged edge-on disks surrounding nearby
Class II objects has been available so far\footnote{While this work
was being revised, \cite{2004ApJ...616..998W} published a
spectroscopic study of a large sample of Taurus-Auriga YSOs,
including the targets studied here. Our long slit spectra complement
these data by providing additional information on the origin of
emission lines in edge-on stars.}. From the observed strong
polarization it is known that only scattered light is reaching us
from such objects. Hence, these objects are much fainter than
ordinary T Tauri stars at the same distance, and large telescopes
are needed to observe them with high resolution optical
spectroscopy.

In order to learn more about the nature and evolutionary stage of
the central objects of these disks and to derive information about
the physical structure and properties of the observed proto-stellar
disk-jet systems, we used the VLT {\sc Uves} spectrograph to study
the resolved edge-on disks surrounding the the central object of
HH30 (in the following referred to as ``HH30$^{*}$'') and the young
stellar objects HV~Tau~C and HK~Tau~B. For comparison we also
included in our {\sc Uves} observing program the prominent classical
T~Tau star (CTTS) HL~Tau and the presumed non-emission PMS star LDN
1551-9.

HH 30$^{*}$, the central object of HH30, was discovered by
\cite{1983ApJ...274L..83M}.  Subsequently HH 30 and HH 30$^{*}$ were
studied by many different authors from the ground
\citep{1987ApJ...319..275M,1990A&A...232...37M,1987ApJ...321..846C,1990PASP..102..972G,
1993ApJ...408L..49R,1998AJ....115.2491K} and with the HST
\citep{1996ApJ...473..437B,1998ApJ...497..404W,2004ApJ...602..860W}.
From a low-resolution optical spectrum \cite{1998AJ....115.2491K}
estimated the spectral type of HH30$^{*}$ to be around M0.

HK~Tau~B was identified as an edge-on disk by
\cite{1998ApJ...502L..65S} from HST and by
\cite{1998ApJ...507L.145K} from ground-based adaptive optics
high-resolution images. A detailed description of the observed
properties of this object can be found in
\cite{2003A&A...400..559D}.

HV~Tau~C was discovered as the third component of the HV~Tau system
by \cite{1992ApJ...384..212S}. After forbidden emission lines were
observed in HV~Tau~C by \cite{1994A&A...287..571M}, HV~Tau~C was
included (as HH 233) in the \cite{Reipurth99} catalog of Herbig-Haro
objects. The true nature of HV~Tau~C was clarified by
\cite{1998A&A...338..122W} who showed that the observed IR flux
could not be explained by an HH nebula but instead required the
presence of a stellar central source. Finally, HV~Tau~C was
identified as an edge-on disk by \cite{2000A&A...356L..75M} using
ground-based AO images and \cite{2003ApJ...589..410S} using HST
imaging.

The comparison object HL~Tau is listed as a CTTS in the PMS star
catalog of Herbig and Rao (1972). Its optical spectrum, showing
strong and broad emission lines superimposed on a veiled late-type
photospheric absorption spectrum corresponding to $\approx $~K7, has
been described, e.g., by \cite{1972ApJ...174..401H},
\cite{1979ApJS...41..743C}, and \cite{1990ApJ...363..654B}.  As in
the case of our program objects, at optical wavelengths we receive
only indirect light that is scattered towards us by the reflection
nebula of HL Tau \citep{1995ApJ...449..888S}, although the central
object is directly visible at IR wavelength
\citep{1997ApJ...478..766C}.  Radio observations of HL Tau have been
interpreted as evidence for the presence of a massive dust and gas
disk observed at an inclination of about $67^\circ$
\citep{1990AJ.....99..924B,1993ApJ...418L..71H,
1997ApJ...478..766C}, surrounded by an extended envelope.  The star
is also known to drive a prominent jet directed approximately
perpendicular to the proposed disk \citep{1990A&A...232...37M,
1996A&A...305..527C}.  All emission lines of HL~Tau are broad and
the forbidden lines are known to show complex, blue-shifted line
profiles \citep{1983RMxAA...7..151A,1994ApJS...93..485H}.

Our spectrum of the comparison object HL~Tau confirms the spectral
properties described in the literature cited above. A detailed
discussion of HL~Tau is outside the scope of the present paper;
however, since our high resolution spectrum provides some improved
information on this object, our new observations will be presented
below together with our results on the edge-on disks.

The second comparison star, LDN~1551-9, located about 7$^\prime$ NE
of HL~Tau, was identified (and classified as a K6 star) during a
search for PMS stars in in the direction of the Lynds Dark Cloud
1551 by \cite{1983AJ.....88..431F}.  Although no line emission was
detected, \cite{2003A&A...403..187F} suggested that LDN 1551-9 is a
PMS star of the Taurus star formation region on the basis of its
X-ray flux. Our {\sc Uves} spectrum of LDN 1551-9 shows a
photospheric absorption spectrum corresponding to about M0 with
narrow emission lines of Ca\,{\sc ii} H and K and an H$\alpha$
absorption line that is partially filled in by emission. However, no
Li\,{\sc i} (1) 6708 \AA\ absorption is detectable (EW $<$ 5 m\AA).
The radial velocity of the star differs by about 10 km s$^{-1}$ from
the CO velocity of LDN 1551 and the typical radial velocity of the
Taurus PMS stars (cf. Table~\ref{RVs}), so that we conclude that LDN
1551-9 is an active late-type star, but very likely not a PMS star.
Nevertheless, its spectral type and the narrow absorption lines,
showing no detectable rotational broadening, make this star well
suited as a comparison star for our program objects.

The paper is organized as follows: Section 2 presents the
observational details; in Section 3 we discuss the observed spectra;
in Section 4 we discuss possible explanations for the observed
spectral properties; and Section 5 presents our conclusions.

\section{Observations and data reduction}

Our data are based on observations carried out in service observing
mode in December 2002 and January 2003 with {\sc Uves}, the {\bf
U}ltraviolet and {\bf V}isual {\bf E}chelle {\bf S}pectrograph at
the Nasmyth platform B of ESO's VLT UT2 (Kueyen) on Cerro Paranal,
Chile. For all observations the standard setting DIC1, 390+580, was
used. The observed wavelength range extended from 3280 to 4490 \AA\
in the blue channel and from 4726 to 6722 \AA\ in the red channel,
except for a gap from 5708 to 5817 \AA\ due to the space between the
two CCDs of the detector mosaic of the red channel. The projected
slit width was 0\farcs8, resulting in a measured FWHM spectral
resolution of about 50\,000.  The median FWHM seeing for the
observations was 0\farcs83, but conditions varied significantly
between the individual frames with extrema of 0\farcs53 and
1\farcs38.

The total integration times (between 5 minutes and 3.9 hours) were
split into individual exposures $< 50$ minutes. Spectra were
obtained for the program objects HH30$^{*}$ and HK~Tau~B with two
different position angles of the projected spectrograph slit, with
the slit oriented either parallel or perpendicular to the disk
plane. In the case of HV~Tau~C only a spectrum with the slit
parallel to the disk was obtained, since with an orientation
perpendicular to the disk the slit would have passed too close to
the bright main component HV~Tau A/B of this triple system. In the
case of the comparison object LDN 1551-9 (expected to be an
unresolved star), the standard NS slit orientation was used. The
slit was oriented parallel to the PA direction 140$^{\circ}$ for the
comparison object HL~Tau, i.e.\ approximately parallel to the
extension of the dust and CO disks reported to be present at the
position of HL Tau and perpendicular to the direction of the HL Tau
jet \citep[see e.g.][]{1997ApJ...478..766C,1990A&A...232...37M}.

\begin{figure}
\resizebox{\hsize}{!}{\includegraphics[angle=0]{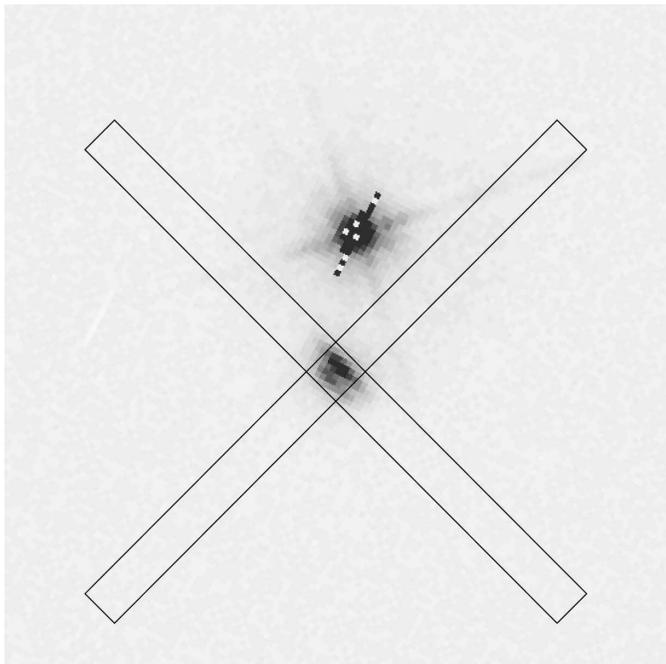}}
\caption[]{HST archive image of HK~Tau~A and B with the positions
and extent of the projected {\sc Uves} red-channel spectrograph
slits used for this study. The slit length is 11\farcs8. North is up
and East to the left.} \label{slits}
\end{figure}

The pixel scales and slit lengths were, respectively, 0\farcs246
pixel$^{-1}$ and 7\farcs6 in the blue and 0\farcs182 pixel$^{-1}$
and 11\farcs8 in the red arm. As an example we present the contours
of the ``red'' slit projected onto an image of HK~Tau~A and B in
Fig.~\ref{slits}, while further information on the observations is
listed in Table~\ref{Observations}.

\begin{table}
\caption{Some technical details of the observed spectra}\label{Observations}
\begin{center}
\small{
\begin{tabular}{lrcr}
\hline
Object & pos.\ angle & nr.\ expos. & tot.\ time (s)\\
\hline
HH 30$^{*}$ & 120$^\circ$ \ ($\parallel$ disk) & 6 & 14\,056 \\
HH 30$^{*}$ & 30$^\circ$ ($\perp $ disk) & 4 & 10\,600 \\
HK Tau B & 135$^\circ$ ($\perp $ disk) & 3 & 8\,460 \\
HK Tau B & 45$^\circ$ \ ($\parallel $ disk) & 3 & 8\,460 \\
HV Tau C & 109$^\circ$ \ ($\parallel $ disk) & 3 & 8\,640 \\
HL Tau & 140$^\circ$ \ ($\parallel $ disk) & 1 & 300 \\
LDN 1551-9 & 0$^\circ$ & 1 & 300 \\
\hline
\end{tabular}
}
\end{center}
\end{table}

The observational data were reduced and two-dimensional spectra
extracted using mostly standard ESO pipeline software for {\sc
Uves}.  An exception was the order-merger procedure where the ESO
software did not produce satisfactory results, mainly because the
very noisy edges of the echelle orders deteriorated the S/N in
overlapping regions; therefore, the order merging was carried out
using software developed at the LSW Heidelberg
\citep{1999oisc.conf..331S}. All spectral frames were converted to
the same (heliocentric) wavelength scale. For the main targets,
where several frames were available, we constructed mean spectra by
calculating the median of the individual spectra.

In addition to the two-dimensional spectra we derived several sets
of one-dimensional spectra by integrating along the direction
perpendicular to dispersion with different integration limits. As
discussed in detail in the following sections, these one-dimensional
spectra were used to obtain information on the central stars and on
various subcomponents of the observed objects. Since all objects are
strongly reddened, the continuum S/N was in all cases $< 1$ at the
blue end of the spectral range, but reached values up to $\approx
60$ at the red limit. Emission lines could generally be detected and
evaluated for $\lambda > 3600$ \AA.

\section{The observed spectra}\label{spectrum}

\subsection{Basic spectral properties}

\begin{table*}
\caption{Observed heliocentric radial velocities of the photospheric
absorption spectra ($v_{*}$), of the H$\alpha $ emission line peaks
($v_{\alpha}$), the He\,{\sc i} emission line peaks
($v_{\mathrm{HeI}}$), the forbidden line peaks
($v_{\mathrm{forbidden}}$), the sharp absorption (or, in the case of
LDN 1551-9, sharp emission) features of the resonance lines of
Na\,{\sc i} and Ca\,{\sc ii} ($v_{\mathrm{Na-D}}$ and
$v_{\mathrm{H\&K}}$), and the velocity of the CO emission from the
corresponding dark clouds ($v_{\mathrm{CO}}$). For the forbidden
lines of HL~Tau the velocities of the blue and the red peaks of the
double-peaked profile are given.  All values are heliocentric
velocities in km s$^{-1}$. Where possible, statistical mean errors
are included. If no error is listed the m.e.\ is about 2 km
s$^{-1}$. }\label{RVs}
\begin{center}
\small{
\begin{tabular}{lccccrccc}
\hline Object & Sp.Type & $v_{*}$ & $v_{\alpha}$  &
$v_{\mathrm{HeI}}$ &
$v_{\mathrm{forbidden}}$ & $v_{\mathrm{Na-D}}$ &  $v_{\mathrm{H\&K}}$ & $v_{\mathrm{CO}}$ \\
\hline HH 30$^{*}$ & K7 & $+21.5 \pm 2.0$  & $+20.3 $ &$+20.4 $ &
$+19.8 \pm 0.3$ & $+20.6 \pm 1.3 $ &  & +19.0 \\
HL Tau & K7  & $+20.7 \pm 2.2$& $+99.6$ &$+31.9$ &
 $ -173.5 \pm 0.5 $ & $+20.7 \pm 1.0$ & $+22.4 $ &$ +19.0 $\\
 &  & & &  &
 $ +15.4 \pm 1.2 $ & & & \\
HK Tau B & M1 & $+17.1 \pm 0.2$  & $+12.5 $ &$+21.2 $ &
$+12.1 \pm 1.0$ & $+18.8 \pm 2.0 $ & $+17.9 $ & +18.0 \\
HV Tau C & M0 & $+20.3 \pm 1.2$  & $+12.5 $ & $+24.7 $ &
$+12.2 \pm 2.4 $ & $+16.8 \pm 0.5 $ & $ +17.5$ & +18.7 \\
LDN 1551-9  & M0  & $+31.0 \pm 0.5 $  &   &   &
   &  & +29.6 & +19.0 \\
\hline
\end{tabular}}
\end{center}
\end{table*}

In our {\sc Uves} spectra all three edge-on disks show similar
general spectral properties. Moreover, apart from the absence or
weakness of permitted metallic emission lines (prominent in HL~Tau),
all our edge-on disks have spectra which qualitatively resemble that
of HL Tau. In particular, all spectra include absorption lines of a
late-type photosphere (with a spectral type around M0) and emission
lines typical of CTTSs. Obviously, all observed edge-on disks
contain CTTSs as central objects. Quantitatively, however, the
edge-on disks do show some distinct differences when compared to
other T~Tauri stars:
\begin{description}
    \item[i)] all three edge-on
objects have exceptionally strong forbidden lines relative to the
continuum flux;
    \item[ii)] as demonstrated in Table~\ref{Linewidths},
the permitted emission lines are narrower than in HL~Tau and other
CTTSs and are (almost) undisplaced relative to the photospheric
spectrum (cf. Table~\ref{RVs}).
\end{description}
As an example we note that the FWHM of the H$\alpha $ emission lines
of our edge-on objects is $< 100$ km s$^{-1}$ in all three cases
(with a mean of $61 \pm 13$ km s$^{-1}$), while for normal CTTS this
value is in the range 100 - 500 km s$^{-1}$ \citep [see e.g.,] []
{1992ApJS...82..247H}. In contrast to many other CTTSs, our edge-on
disks show no, or only very weak, displaced or broad forbidden-line
components, and all emission lines exhibit simpler and more
symmetric profiles than normally observed in CTTSs.

As shown by Fig. 2, the spectra of the three edge-on objects are
veiled. The amount of veiling -- defined by the ratio {\it [excess
flux]/[photospheric flux]} and estimated from the rest intensities
of the strongest photospheric absorption lines and by comparing the
line strengths with those observed in LDN 1551-9 in the red spectral
range (around 6400 \AA ) -- is about 0.2 for HK Tau B and
approximately 0.5 for HH30$^{*}$ and HV Tau C. The different amounts
of veiling may indicate that the central objects of our targets
cover a range of PMS evolutionary stages or at least a range of
accretion rates.

For all three edge-on objects the veiling is weaker than in HL~Tau,
which again indicates either more modest disk accretion rates or a
more advanced evolutionary stage of the central stars. Given the
moderate veilings of our target stars, the large equivalent widths
of their emission lines (Table~\ref{EWs}) is unexpected.

\begin{figure}
\resizebox{\hsize}{!}{\includegraphics[angle=-90]{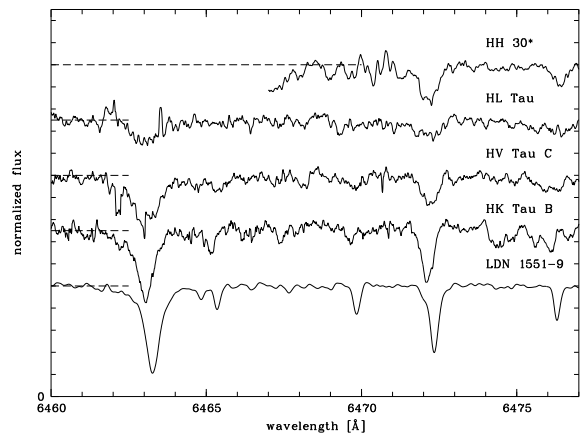}}
\caption[]{Section of the photospheric spectrum of two of the
edge-on disks and of the two comparison objects. The four spectra
have been normalized to the same continuum level of 1.0. To avoid
overlapping, the zero levels have been shifted vertically by 0.5,
1.0, 1.5, and 2.0, respectively, for HK~Tau~B, HV~Tau~C, HL~Tau, and
HH 30$^{*}$. The 6460 - 6467 \AA \ section of the HH 30 spectrum was
affected by instrumental defects and therefore omitted.

 }
\label{phot_spec}
\end{figure}

\begin{table}
\caption{Observed average FWHM (in km s$^{-1}$) of the observed
emission lines (except for the complex forbidden lines of HL~Tau,
for which we give the full width at zero intensity). If no error
is listed the m.e.\ is $\leq $ 10\% of the measured value.
}\label{Linewidths}
\begin{center}
\small{
\begin{tabular}{lrrrrr}
\hline
Object & H$\alpha $ & He\,{\sc i} & Na\,{\sc i} (D) & Ca\,{\sc ii} H\&K & Forb. Lines \\
\hline
HH 30$^{*}$ & 50 & 35 & 33 & ? & $ 25 \pm 1 $\\
HL Tau & 267 & 170 & 149 & 130 & $\approx $ 300\\
HK Tau B & 37 & 34 & 34 & 25 & $ 33 \pm 3 $ \\
HV Tau C & 96 & 58 & 36 & 43 & $ 51 \pm 1 $ \\
\hline
\end{tabular}
}
\end{center}
\end{table}

\begin{table}
\caption{Observed equivalent widths (in $\AA$) of selected strong
emission lines. }\label{EWs}
\begin{center}
\small{
\begin{tabular}{lrrr}
\hline
Object & H$\alpha $ & [O\,{\sc i}]$\lambda 6364$ & [S\,{\sc ii}]$\lambda 6716$\\
\hline
HH 30$^{*}$ & 454$\pm$15 & 100.2$\pm$1.5 & 104.5$\pm$0.2 \\
HL Tau & 78$\pm$0.2 & 1.3$\pm$0.2 & 4.5$\pm$0.2 \\
HK Tau B & 9.6$\pm$0.2 & 0.4$\pm$0.1 & 0.9$\pm$0.2 \\
HV Tau C & 49.2$\pm$0.2 & 7.8$\pm$0.1 & 6.3$\pm$0.1 \\
\hline
\end{tabular}
}
\end{center}
\end{table}

\subsection{The photospheric absorption spectra}

As already noted, all our edge-on disk spectra contain absorption
lines with line profiles and relative line strengths that agree well
with those of late-type stellar photospheres. From comparing the
relative line strengths with MK standard star spectra observed at a
similar spectral resolution and by evaluating the shapes (but not
the absolute strength,  which may be affected by veiling) of the
TiO-Bands, we estimated the approximate spectral types listed in
Table~\ref{RVs}.  The luminosity classes appear to be about IV but
cannot be determined precisely from our spectra since they show
veiling effects and since we have a sufficient continuum S/N only in
the red where few diagnostic lines are available. Because of the
limited wavelength range usable for classification, the spectral
types listed in Table~\ref{RVs} cannot be regarded as more accurate
than those derived from lower resolution spectra.

\begin{figure}
\resizebox{\hsize}{!}{\includegraphics[angle=-90]{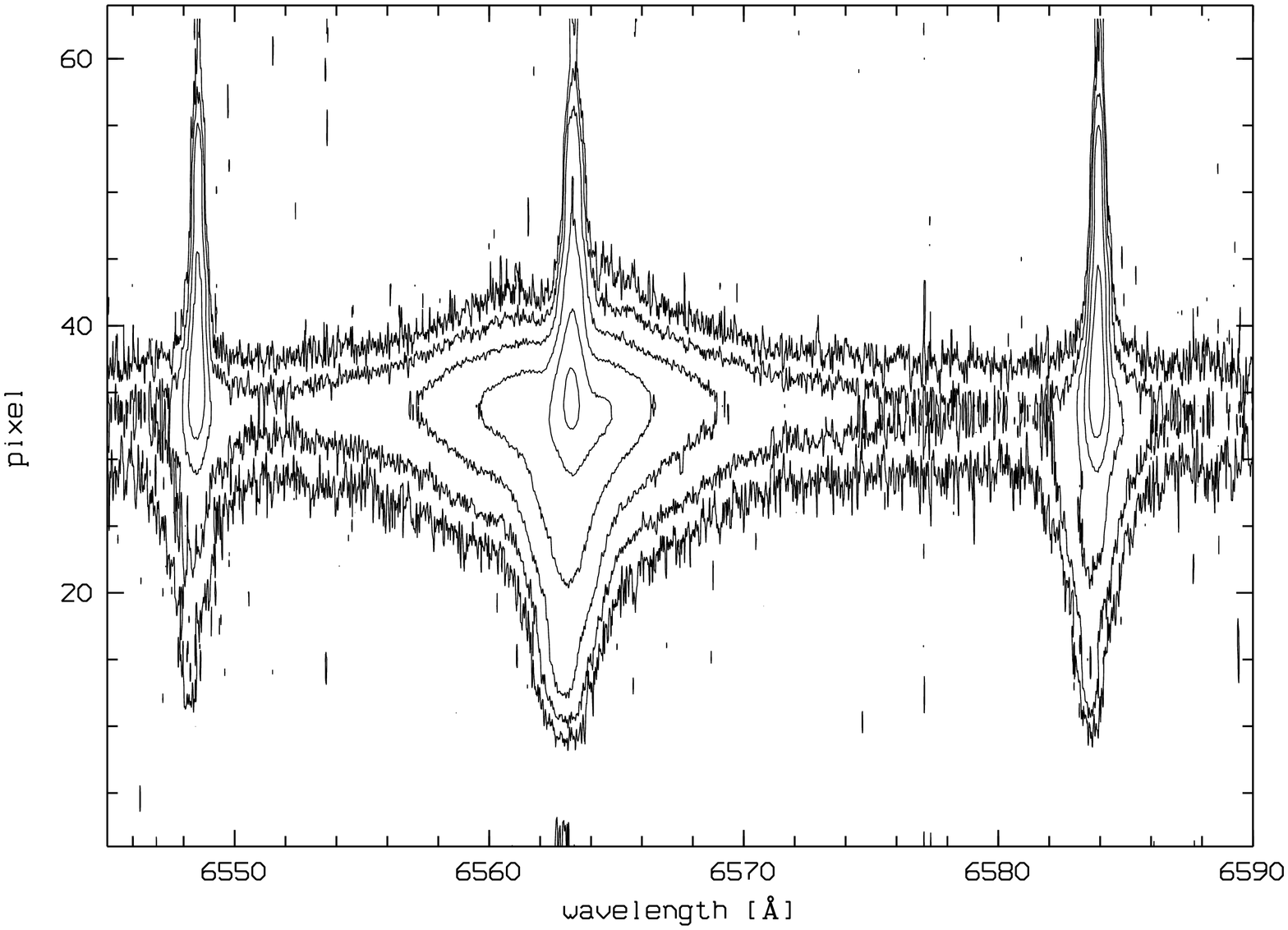}}
\resizebox{\hsize}{!}{\includegraphics[angle=-90]{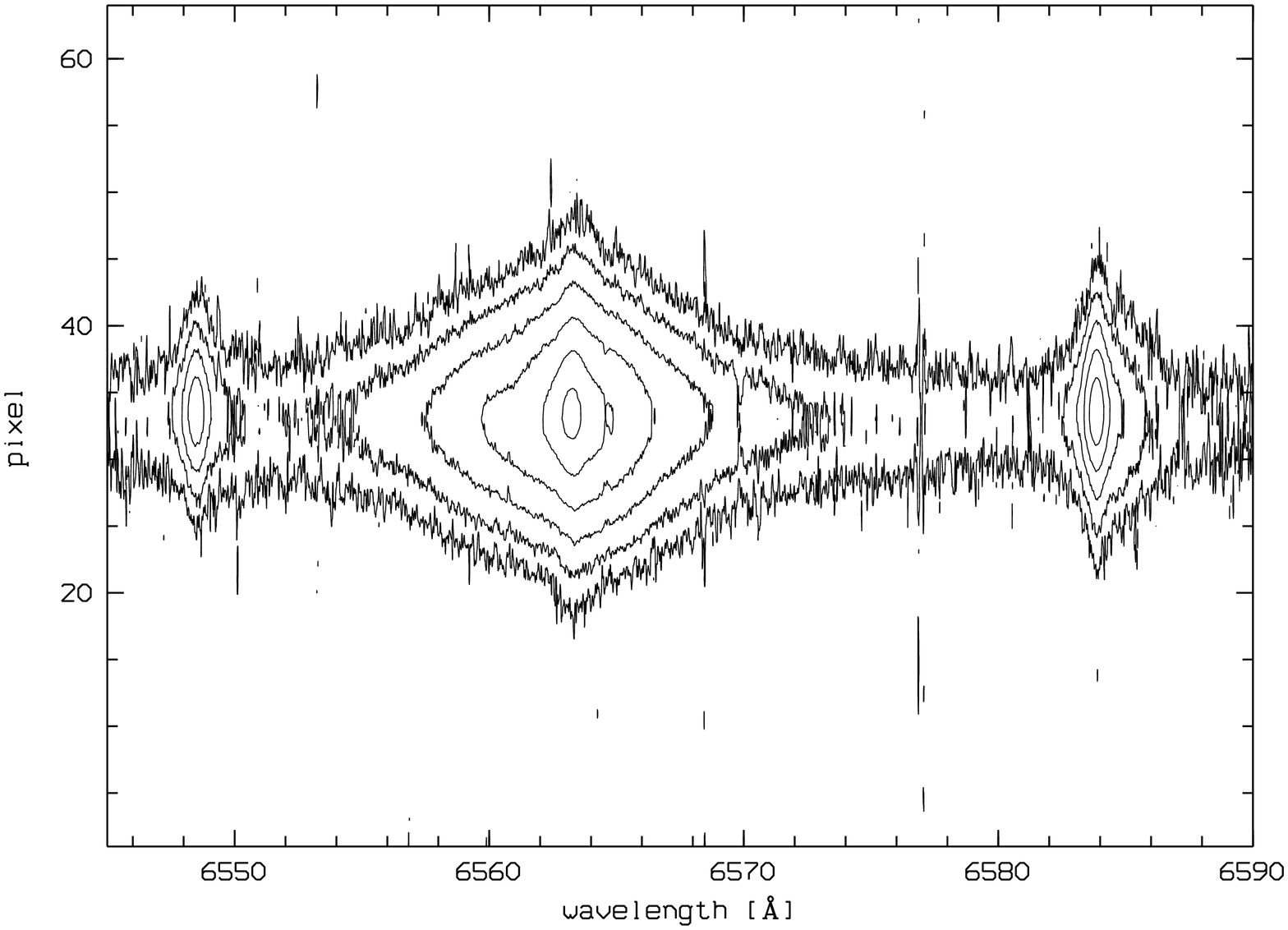}}
\caption[]{Contour diagrams of the 2-D spectra of HH30$^{*}$ (a)
with the slit oriented parallel to the jet, and (b) with the slit
parallel to the disk, showing the H$\alpha $ and the adjacent
[N\,{\sc ii}] lines. Between two contours the intensity changes by a
factor of 3.16 (= $\sqrt{10}$).
 }
\label{hh30_ha_disk}
\end{figure}

As again illustrated by Fig.~\ref{phot_spec}, the photospheric line
profiles of all our three edge-on objects are narrow. In the case of
HK~Tau~B, the low veiling allows us to derive an upper limit for a
possible rotational broadening of $v \times \sin i < $ 10 km
s$^{-1}$ from the weak photospheric lines. The profiles observed in
the spectra of HH30$^{*}$ and HV~Tau~C are consistent with such low
$v \times \sin i$ values as well. But, since only the {\it
intrinsically} broad strong lines could be detected in these
spectra, the observational upper limit for rotational broadening is
as high as $20$ km s$^{-1}$. These limits demonstrate that no
rotation  is observed to be higher than normally expected for CTTSs,
which gives further support to the normal CTTS character of the
central objects of the edge-on disks.

\subsection{Wind and jet emission lines}

In T~Tauri stars the winds and outflows, which often are collimated
towards the rotation axes and then called ``jets'', are known to
produce extended emission regions dominated by low-ionization
forbidden emission lines. In the spectra of all our three edge-on
disks and in the spectrum of HL~Tau, all expected lines of [N\,{\sc
ii}], [O\,{\sc i}], [S\,{\sc ii}], and [Fe\,{\sc ii}] were clearly
detected. HH30$^{*}$, HV~Tau~C and HL~Tau also showed [N\,{\sc
i}]$\lambda\lambda$ 5198, 5200 emission while  [O\,{\sc
ii}]$\lambda\lambda$3726, 3729 and [O\,{\sc i}]$\lambda$5577 could
be reliably detected only in HH30$^{*}$ and HV~Tau~C.  The relative
strength of the [O\,{\sc ii}] doublet components indicate rather
high electron densities for the volume producing these lines (n$_{e}
> 10^{5}$ for HH30$^{*}$ and $\approx 10^{4}$ for HV~Tau~C).

\begin{figure}
\resizebox{\hsize}{!}{\includegraphics[angle=-90]{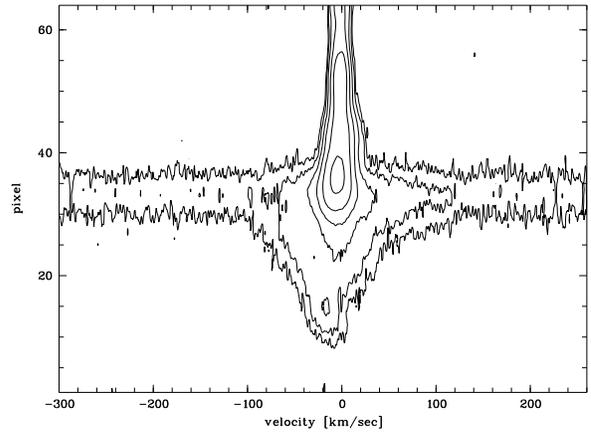}}
\caption[]{Contour diagrams of the 2-D spectrum of HH30$^{*}$
showing the [S\,{\sc ii}] 6717 \AA\ line with the slit oriented
parallel to the jet. Between two contours the intensity changes by a
factor of 3.16 (= $\sqrt{10}$). The zero-point of the velocity scale
corresponds to the object's adopted systemic velocity of +20.1 km
s$^{-1}$.} \label{hh30_jet}
\end{figure}

As illustrated by Figs.~\ref{hh30_ha_disk},~\ref{hh30_jet},
and~\ref{hktau_sii}, extended jet emission was observed in the two
spectra which were taken with the spectrograph slit oriented
perpendicular to the (projected) disk plane (i.e.\ the spectra
PA=30$^{0}$ of HH30$^{*}$ and PA=135$^{0}$ of HK~Tau~B).  On the
other hand, extended forbidden-line emission was {\it not}
\/detected in any of the other spectra, which all were taken with
the slit parallel to the (projected) disk planes. In fact, the
forbidden-line emission region was unresolved, i.e.\ the FWHM extent
of the emission region was not larger than that of the seeing
profile, in all our edge-on disk spectra taken with a disk-parallel
slit. This finding is consistent with the result of
\cite{1996ApJ...473..437B}, who found for the HH30 jet a width at
the base $\le 20$ AU, which is below the resolution of our
ground-based observations.  The continuum emission in the
disk-parallel spectra always showed an extension significantly
larger than the forbidden-line emission. Obviously, most the
forbidden-line flux that entered the spectrograph slit reached us
directly from an unresolved emission region above the disk while the
continuum emitted by the central star is scattered towards us by an
extended and resolved dusty region.

As pointed out (e.g.) by \cite{2003A&A...397..675T} line emission
regions slightly off-set from the center of the continuum emission
can in theory also be produced by unresolved, very close binary
components. However, for our edge-on object a jet origin of the
observed forbidden-line emission appears much more plausible.

For each of the three edge-on program objects all forbidden lines
(including the [Fe\,{\sc ii}] lines) were found to show identical or
qualitatively similar line profiles in the one-dimensional mean
spectra averaged over the two slit orientations. On the other hand,
these profiles differ between the different targets.

\subsubsection{HH30$^{*}$}

\begin{figure}
\resizebox{\hsize}{!}{\includegraphics[angle=-90]{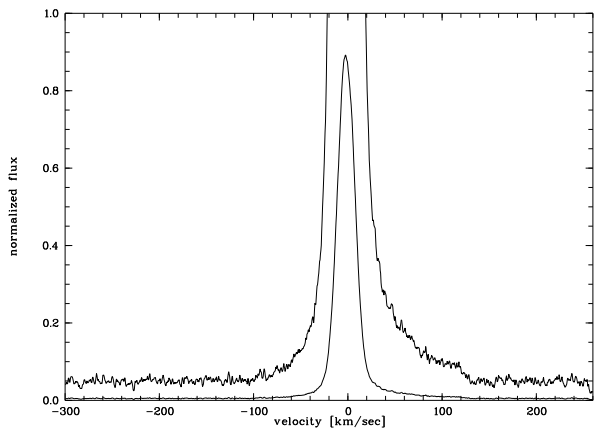}}
\caption[]{Profile of the [S\,{\sc ii}] 6717 \AA\ line of
HH30$^{*}$, plotted at two different vertical scales (differing by a
factor 10) to show the two different components. The zero-point of
the velocity scale corresponds to the object's systemic velocity.
 }
\label{hh30_sii}
\end{figure}

In the case of HH30$^{*}$ the forbidden-line profiles are dominated
by narrow cores with a heliocentric peak radial velocity of $+19.8
\pm 0.3$ km s$^{-1}$. This value is (within the error limits) in
agreement with the weighted mean of all velocities measured in the
integrated spectrum of HH30$^{*}$ ($20.1 $ km s$^{-1}$), which was
adopted as the systemic velocity of HH30$^{*}$.  Underlying the
narrow cores, the forbidden-line profiles include much weaker and
redshifted broad components (FWHM $>$ 100 km s$^{-1}$, see
Fig.~\ref{hh30_sii}).

The resolved extended emission in the direction of the prominent
HH30 jet \citep[i.e.\ towards PA$\approx $30$^{0}$,
cf.][]{1996ApJ...473..437B} shows only the narrow profile
components, which are slightly redshifted by about 3.0 km s$^{-1}$
(see Figs.~\ref{hh30_ha_disk} \&~\ref{hh30_jet}). The weaker and
less extended emission towards PA$\approx $210$^{0}$ (usually
referred to as the ``counter-jet'') shows somewhat broader line
profiles that are blue-shifted by about $13.0 $ km s$^{-1}$ relative
to our adopted systemic velocity.  If these small velocity shifts
are due to a tilt of the jet, they would indicate that -- contrary
to the expectation from the morphology -- the counter jet is
approaching us, while the north-northeast emission region is
receding. However, the small differences in the systemic velocity
show that any tilt, if present, is very small. Since the HH30 jet is
known to be slightly curved \citep[cf.][]{1996ApJ...473..437B}, the
observed velocity shifts may simply reflect small deviations from a
linear flow.

\begin{figure}
\resizebox{\hsize}{!}{\includegraphics[angle=0]{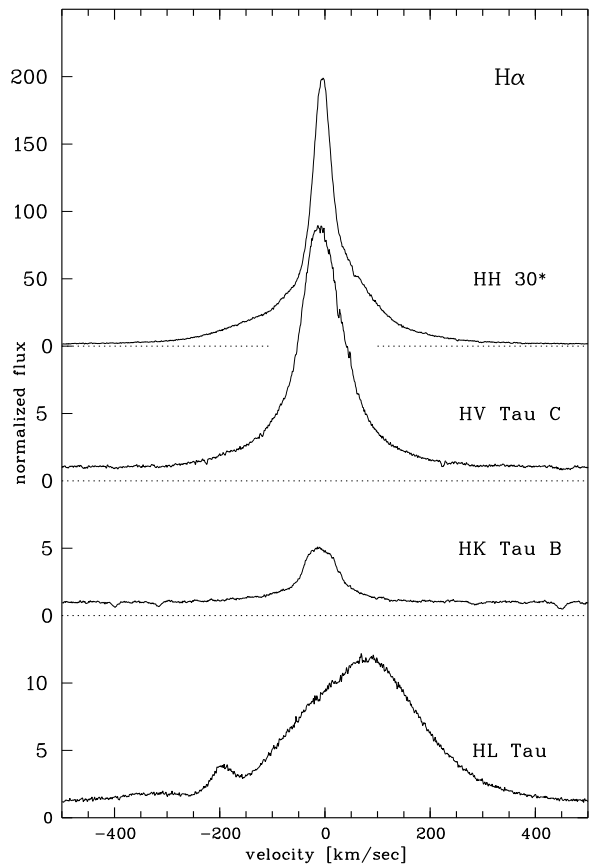}}
\caption[]{Profile of the H$\alpha $ line of our three targets and
of the comparison CTTS HL Tau. The zero-point of the velocity scale
corresponds to the objects' systemic velocity. The sharp absorption
at about $+50$~km s$^{-1}$ is an artefact. } \label{hh30_ha_prof}
\end{figure}

\begin{figure}
\resizebox{\hsize}{!}{\includegraphics[angle=-90]{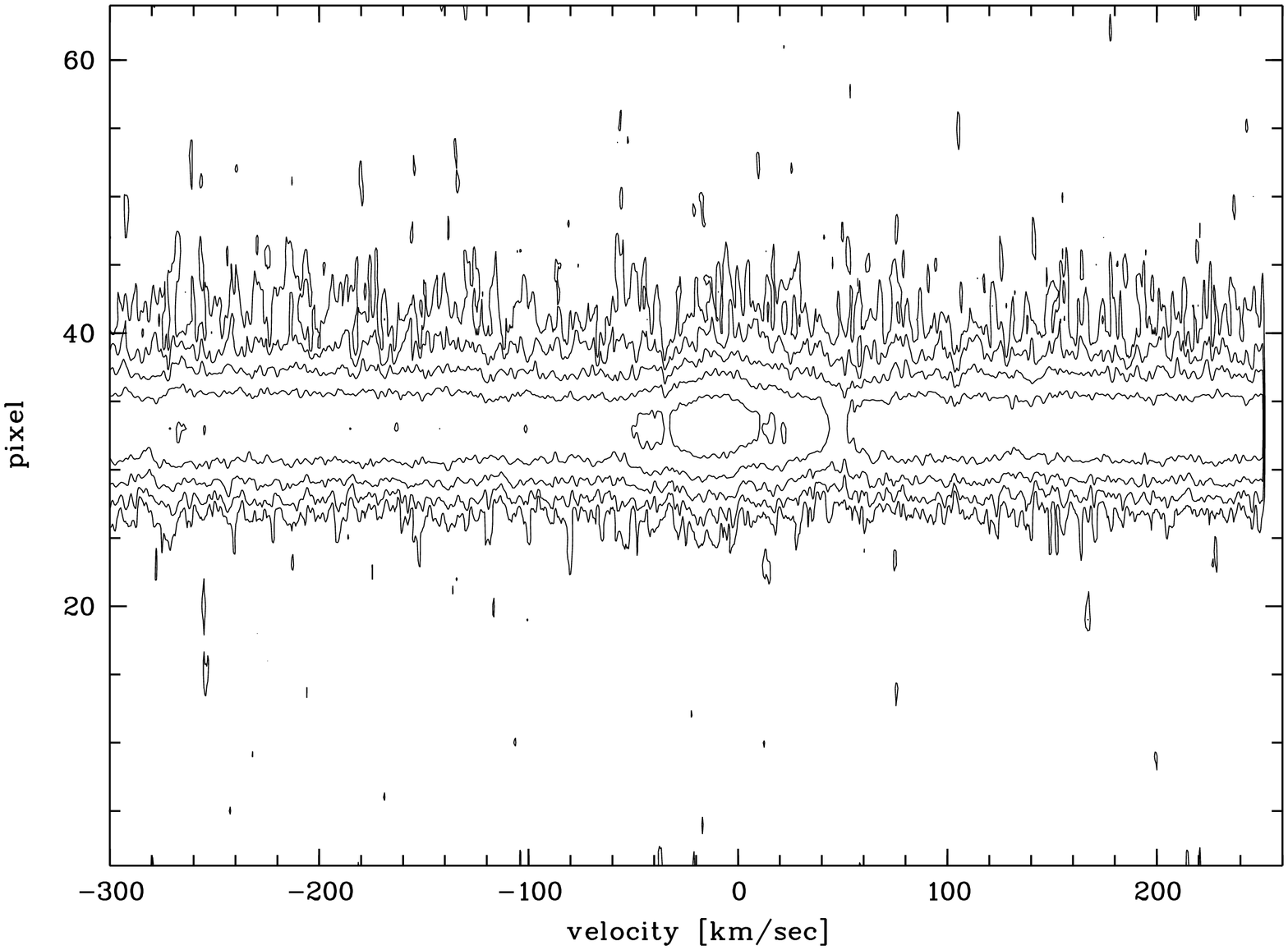}}
\resizebox{\hsize}{!}{\includegraphics[angle=-90]{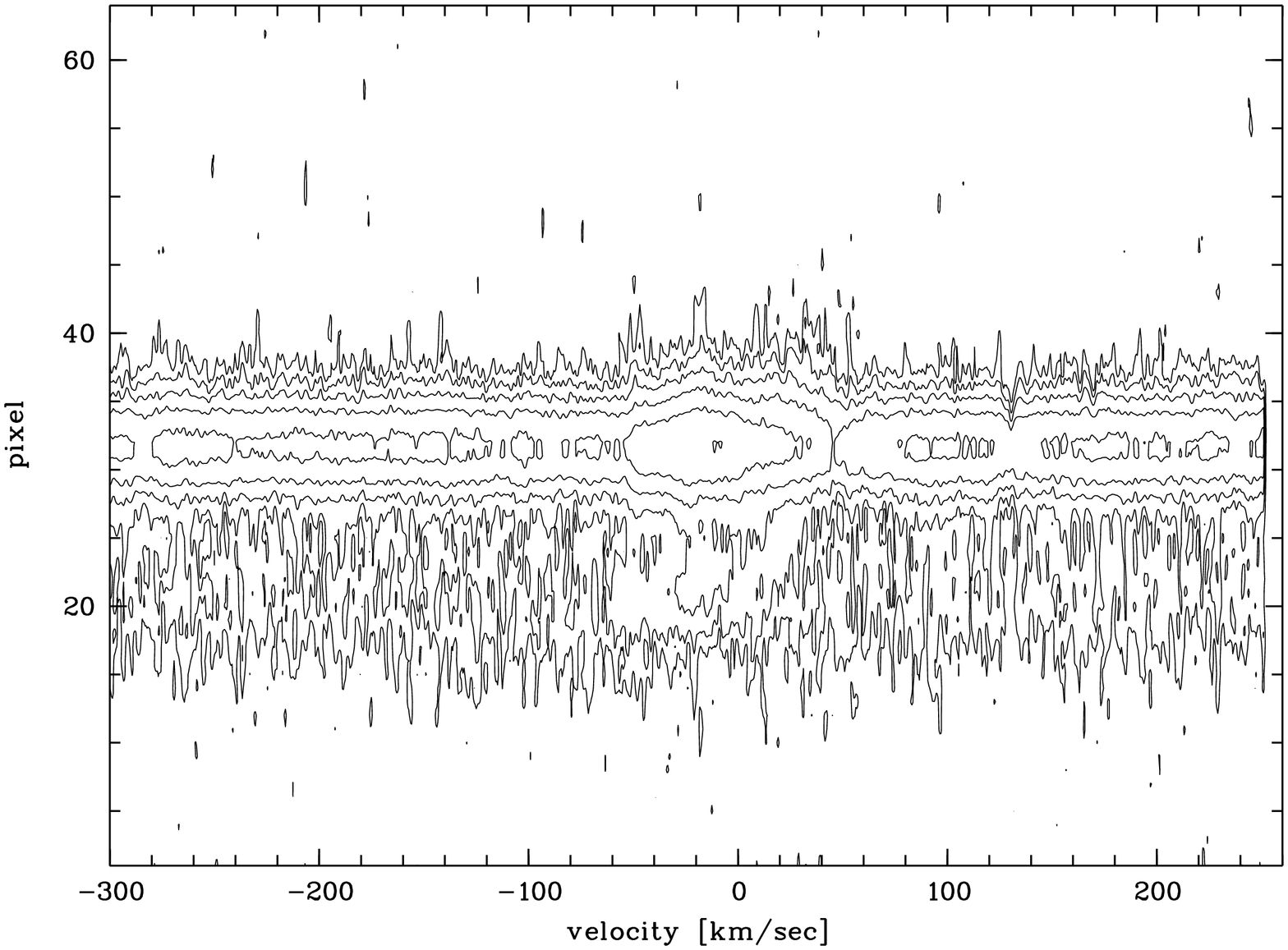}}
\caption[]{Contour diagrams of the 2-D spectra of HK~Tau~B showing
the region of the [S\,{\sc ii}] 6717 \AA\ line, (a) with the slit
oriented parallel to the jet, and (b) with the slit parallel to the
disk. Between two contours the intensity changes by a factor of 2.
The zero-point of the velocity scale corresponds to the object's
systemic velocity.
 }
\label{hktau_sii}
\end{figure}

The narrow (jet) line profile component is also visible in H$\alpha
$ (cf. Fig.~\ref{hh30_ha_disk} and~\ref{hh30_ha_prof}) and in the
He\,{\sc i} emission lines. As illustrated by
Fig.~\ref{hh30_ha_disk} (and Fig.~\ref{hh30_ha_prof}), the H$\alpha$
line profile also includes a very broad component (extending to at
least $\pm$ 340 km s$^{-1}$ from the line center) at the position of
the continuum source (but not in the jet). Moreover, in the area
covered by our spectrograph slit the jet and the counter jet show
nearly the same H$\alpha $ flux, while the forbidden-line flux is
very different.

\subsubsection{HK~Tau~B}

As illustrated by Fig.~\ref{hktau_sii}, our HK~Tau~B spectrum, taken
with the slit perpendicular to the disk, shows extended
forbidden-line emission directed towards PA 315$^\circ$. In this
direction the slit passes unfortunately close to the position of the
brighter star HK~Tau~A, which is known to show exceptionally strong
H$\alpha$ emission \citep{1998AJ....115.2491K}. Therefore, the
H$\alpha$ emission from the HK~Tau~B jet is contaminated by
straylight from the H$\alpha$ line of HK~Tau~A.  HK~Tau~A may also
contribute to the observed forbidden line flux observed below the
disk of HK~Tau~B, but the observed intensity distribution
perpendicular to the dispersion direction rules out any possibility
that all the extended emission is due to straylight from HK~Tau~A.
The observed jet emission of HK~Tau~B is again dominated by a narrow
and almost unshifted ($\Delta v \leq $ 5 km s$^{-1}$) line
component, while an underlying broader component could not be
detected. However, since the forbidden lines are weaker in HK~Tau~B
and since the photospheric spectrum (less veiled) is more prominent,
a broad component as weak as in the case of HH30$^{*}$ would not be
detectable in our HK~Tau~B spectra.

\subsubsection{HV~Tau~C}

\begin{figure}
\resizebox{\hsize}{!}{\includegraphics[angle=-90]{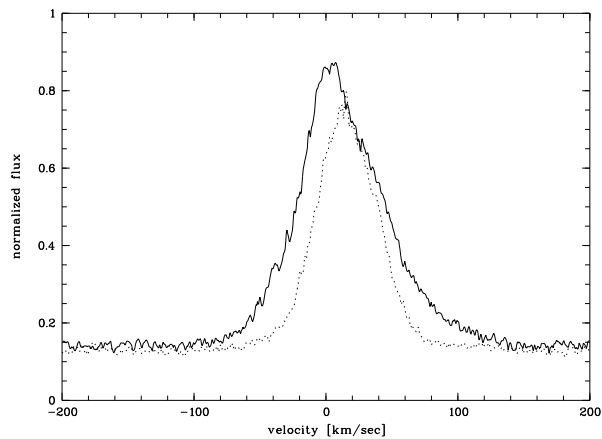}}
\caption[]{Line profiles of the [N\,{\sc ii}] 6583 (solid line) and
[S\,{\sc ii}] 6717 \AA\ lines in the spectrum of HV~Tau~C. The
zero-point of the velocity scale corresponds to the object's adopted
systemic velocity.} \label{hvtau_forbidden}
\end{figure}

In HV~Tau~C the forbidden lines are somewhat broader than in
HH30$^{*}$ and HK~Tau~B (see Table~\ref{Linewidths}), and some line
profiles appear slightly asymmetric, with the red wing slightly more
extended than the blue wing. A slight asymmetry is most pronounced
at the [N\,{\sc ii}] lines, while [S\,{\sc ii}]$\lambda$6716 is
practically symmetric. As illustrated by Fig.~\ref{hvtau_forbidden},
the base of all forbidden lines of HV~Tau~C has about the same
redshift. Hence the asymmetry appears to be caused by a shift of the
emission peak towards shorter wavelengths.  A possible explanation
of the asymmetry could be the presence of two profile components, as
observed in HL~Tau (Fig.~\ref{hltau_forbidden}), but with a velocity
separation that is small compared to the line width in the case of
the (almost) edge-on disk object HV~Tau~C.

\subsubsection{HL~Tau}

\begin{figure}
\resizebox{\hsize}{!}{\includegraphics[angle=-90]{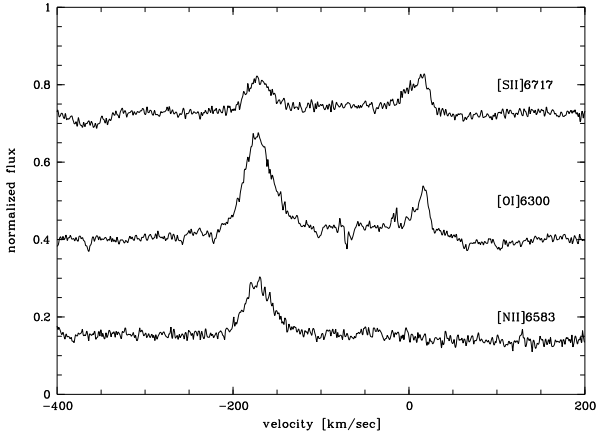}}
\caption[]{Line profiles of the [N\,{\sc ii}] 6583,[O\,{\sc i}]
6300, and[S\,{\sc ii}] 6717 lines in the spectrum of HL~Tau. The
abscissa gives the heliocentric radial velocity}
\label{hltau_forbidden}
\end{figure}

Our (presumably non-edge-on) comparison object, HL~Tau, has been
known to have blue-shifted and, in some cases, double-peaked
forbidden-line profiles
\citep{1983RMxAA...7..151A,1994ApJS...93..485H,1997A&AS..126..437H}.
This is confirmed by our new observations with improved accuracy. As
shown by Fig.~\ref{hltau_forbidden} and Table~\ref{RVs}, one
forbidden line component is blue-shifted by 194 km s$^{-1}$ relative
to the photospheric spectrum, while the other component shows almost
the same radial velocity as both the absorption spectrum and the
sharp resonance line absorption components of HL~Tau.
\cite{1984A&A...141..108A} and others pointed out that such profiles
are not unexpected for disk-jet systems seen at intermediate
inclinations. As illustrated by Fig.~\ref{hltau_forbidden}, the
relative strength of the two profile components varies greatly
between the different forbidden lines, indicating that the two
components originate in different volumes, \citep[as also observed
in some other CTTSs, see][]{1997A&AS..126..437H}. This is confirmed
by our two-dimensional HL~Tau spectrum (taken with the slit parallel
to the projected disk), where the emission region producing the
blue-shifted peak is unresolved, and thus probably originates in a
directly observed narrow jet intersected by the slit, while the
region producing the red peak is extended and shows about the same
angular extension as the continuum.

It seems plausible that the low-velocity emission originates in a
rarefied extended emission region, as proposed for other CTTSs with
similar profiles by \cite{1988ApJ...332L..41K}.  The slight
blue-shift ($\approx $ 5 km s$^{-1}$ relative to the photospheric
spectrum) could be caused either by a slow outflow or by scattering
in a slowly moving medium. That the two components of the [S {\sc
ii}] line are of similar strength in Fig.~\ref{hltau_forbidden},
while the [S {\sc ii}]$\lambda$6731 profile published by
\cite{1997A&AS..126..437H} is dominated by a much stronger
high-velocity component is probably due to our slit orientation
approximately perpendicular to the jet, which suppresses much of the
jet emission.

\subsection{Circumstellar spectral features}

The CTTS disks are known to be essentially composed of relatively
dense cool and dusty gas, so we expected to see spectroscopic
signatures for the disks mainly in the profiles of the
low-ionization resonance doublets of Ca\,{\sc ii} (H \& K) and
Na\,{\sc i} (D). The observed features were in all cases well
separated from the corresponding telluric lines because of the high
resolution of the {\sc Uves} spectra and the significant systemic
radial velocities of all observed objects.

\begin{figure}
\resizebox{\hsize}{!}{\includegraphics[angle=0]{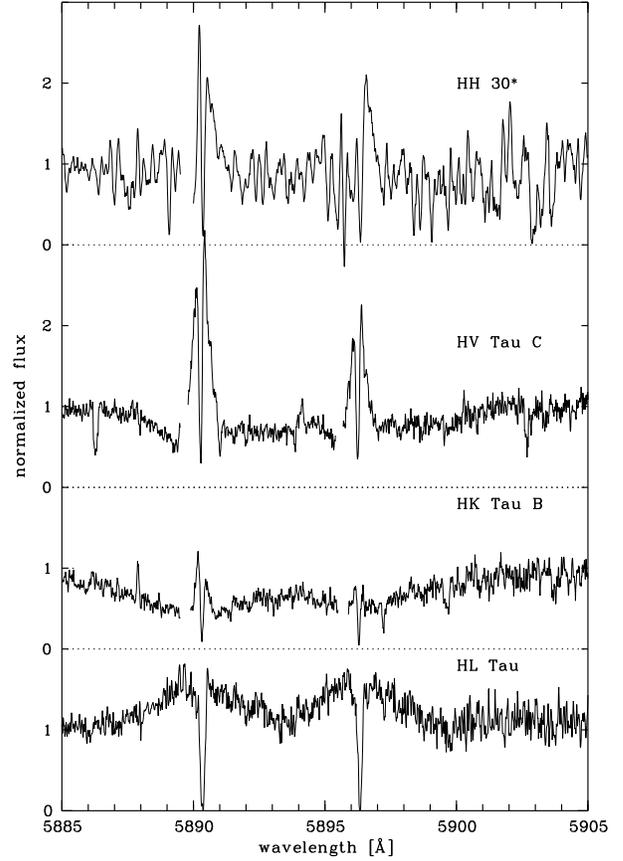}}
\caption[]{Na\,{\sc i} (D) resonance line profiles in the spectra of
HH 30$^{*}$, HV~Tau~C, HK~Tau~B, and HL~Tau.} \label{profile_nad}
\end{figure}

All three edge-on objects and HL~Tau show the Ca\,{\sc ii} and
Na\,{\sc i} resonance doublets in emission, although the lines are
surprisingly weak in HH30$^{*}$. In the edge-on disks the resonance
line emission profiles are generally narrow with FWHM values similar
to those of the forbidden lines and the He\,{\sc i} emission lines
(cf. Table~\ref{Linewidths} and Fig.~\ref{profile_nad}).  An
exception may be Ca\,{\sc ii} H \& K of HH30$^{*}$, where weak
broader emission lines appear to be present. However, the S/N of our
spectra did not allow us to reliably measure the widths of these
very weak features. The non-edge-on system HL~Tau shows much broader
and stronger (relative to the continuum) Ca\,{\sc ii} and Na\,{\sc
i} resonance emission. In Fig.~\ref{profile_nad} we compare the
Na\,{\sc i} (1) emission profiles of the edge-on disks with those of
HL~Tau.  The region from which the resonance line emission is
received has about the same angular extension as the continuum in
all objects, so that the directly visible parts of the jets do not
seem to contribute to resonance line emission. Most of the Na\,{\sc
i} emission must originate near the star, whether in the central
region of the disk or at the base of the jet, and is (like the
continuum) obviously scattered towards us by matter above and below
the disk plane. A more detailed discussion of the origin of this
emission will be given in Section 4.

In addition to the emission we found at all Na\,{\sc i} lines, and
at all Ca\,{\sc ii} resonance lines observed with sufficient S/N,
very narrow absorption components with radial velocities agreeing
well with those of the photospheric spectra (cf. Table 2). For all
three edge-on objects the profiles of these absorption components do
not deviate significantly from the instrumental profile (FWHM
$\approx $ 6 km s$^{-1}$). In the case of HL~Tau the width of the
absorption component corresponds to about twice the instrumental
profile. For HH 30$^{*}$, HK Tau B (and HL Tau) the narrow Na\,{\sc
i} absorption components reach well below the level of the adjacent
continuum and almost to zero intensity, indicating that the
absorption affects practically all light reaching us at this
wavelength. Hence, the absorption must at least in part be of
interstellar or circumstellar origin. Because of the absorption
strength and the excellent agreement of the observed velocities with
the objects' systemic velocities, a circumstellar origin appears
most plausible. Absorption could, in fact, take place in the same
cool dusty matter that scatters the light from the central objects
and the inner disks towards us.

In case the absorbing matter were related to the disks, our 2-D
spectra taken with the slit parallel to the disk might well contain
information on the disk rotation. If the absorbing matter were
orbiting at 100 AU around a central object of one solar mass we
would expect an orbital velocity of about 3 km~s$^{-1}$, which is
within reach of {\sc Uves} spectra. Therefore, we checked our 2-D
spectra for a possible tilt of the sharp absorption component, but
we found no evidence for rotation. The most accurate measurement
(Na\,{\sc i}\,D of HV~Tau~C) resulted in a (non-significant) tilt of
0.48 $\pm $ 0.26 km~s$^{-1}$ arcsec$^{-1}$.  Since seeing tends to
smear possible rotational signatures in our ground-based spectra,
our result does not rule out rotational disks velocities of a few
km~s$^{-1}$ as estimated above.

\section{Interpretation and discussion}

\subsection{Forbidden line emission}
As pointed out above, in all three edge-on objects there is clear
evidence that the narrow forbidden line components are emitted by
jet flows that are seen directly above and below the disks. The
small line width of the narrow forbidden-line components and the
small velocity shifts are readily explained by the fact that in the
edge-on objects the gas in the jet and counter jet is moving
perpendicular to the line-of-sight within a few degrees. The weaker,
broad, and shifted components observed from the same extended area
from which we observe the continuum flux can possibly be explained
by additional light from the jets scattered towards us above and
below the disk by the same material that scatters the light from the
central star.  As the jet plasma moves supersonically with a
significant line-of-sight velocity component relative to the
scattering medium, with most of the jet material moving away from
the scattering layer, such radiation is expected to show significant
line broadening and a velocity shift, as was indeed observed.

The weakness of the forbidden-line emission from the counter-jet of
HH 30$^{*}$, in spite of strong H$\alpha $ emission, may indicate
that this radiation is produced in a relatively dense medium. Either
the counter jet is moving into a much denser environment, which
could also explain the smaller extent, or we see in H$\alpha $ light
coming from a different emission region that is scattered in our
direction.

\subsection{Wind contribution to H$\alpha $ and He\,{\sc i} lines}
Current models for CTTSs attribute H$\alpha $ emission to the
magnetospheric accretion region in all but the most extreme CTTSs
\citep{1998AJ....116..455M, 2001ApJ...550..944M}. Before the
magnetospheric accretion paradigm became consensual, a number of
other possible origins for Balmer line emission had been discussed
in the literature, for example, strong winds
\citep[e.g.,][]{1964ApJ...140.1409K, 1982ApJ...261..279H}, deep
chromospheres \citep[e.g.,][]{1984ApJ...277..725C}, or disk boundary
layers \citep{1989ApJ...341..340B}. Because they can form in a large
range of conditions, it is particularly difficult to disentangle the
contributions to the Balmer lines from all possible emission
regions. As we have seen above, the particular geometry of edge-on
objects allows us to isolate the contribution of the jet to
forbidden emission. The H$\alpha $ and He\,{\sc i} lines share many
similarities with the forbidden lines in edge-on objects, and can be
understood in the same way. In particular, the narrow H$\alpha $ and
He\,{\sc i} emission cores can be understood as jet emission seen
above and below the disk, so they provide direct evidence that the
jet contributes to the formation of strong permitted emission lines
in these objects.

Since we have noted above the strong similarities between edge-on
stars and other CTTSs, the finding that the jet contributes to
H$\alpha$ suggests that emission from the CTTS outflow is an
essential ingredient of H$\alpha$ emission even in moderately veiled
CTTSs. This contrasts with current magnetospheric accretion models,
and suggests that the view angle under which a given CTTS is seen
partly determines the properties of its emission lines.

In order to test this new hypothesis, we now investigate a
cross-section of Taurus-Auriga CTTSs for which (a) precise
photometric periods and radial velocities and (b) simultaneous
determinations of H$\alpha$ equivalent widths and veiling are known.
This sample is given in Table~\ref{OTTSs}, where the star name and
Herbig and Bell catalog number are given in Columns 1 and 2. The
photometric periods and radial velocity measurements, originating
from a compilation of the literature by J. Bouvier (priv. comm.),
are given in Column 3 and 4 respectively. The photospheric radius,
derived as in \citet{1989AJ.....97.1451S}, is listed in Column 5.
The view angle inferred from the previous quantities is given in
Column 6, while Columns 7 and 8 list the H$\alpha$ equivalent widths
and veiling values in the red spectral range, as determined by
\citet{2000AJ....119.1881A} from high-resolution Hamilton
spectrograms. To these 12 stars, we add the 3 edge-on TTSs, for
which the inclination angle is close to 90\deg. Their EWs are given
in Table~\ref{EWs}.

\begin{table}
\caption{Properties of Taurus-Auriga CTTSs with known inclination
angles and simultaneous EW(H$\alpha$) and veiling determinations.
Details and references are given in the text.} \label{OTTSs}
\begin{center}
\small{\begin{tabular}{lcccccccc}

  \hline\hline
  (1) & (2) & (3) & (4) & (5) & (6) & (7) & (8)\\
  Star & HBC & P$_{\rm rot}$ & \emph{v} sin \emph{i} & R & $i$ & EW(H$\alpha$) & v\\
   & & [d] & [km/s] & [R$_\odot$] & [$^{\rm o}$] & [$\AA$] &\\
  \hline
  BP Tau & 32 & 7.6 & 7.8 & 2.2 & 32 & 48.6 & 0.5\\
  DE Tau & 33 & 7.6 & 10  & 1.78 & 57 & 66.5 & 0.9\\
  T Tau & 35 & 2.8 & 20.1 & 4.2 & 15 & 63.1 & 0.1\\
  DF Tau & 36 & 8.5 & 16.1 & 3.45 & 50 & 61 & 1.8\\
  DG Tau & 37 & 6.3 & 21.7 & 2.8 & 58 & 60.5 & 2\\
  DK Tau & 45 & 8.4 & 11.4 & 2.7 & 44 & 16.7 & 0.4\\
  GG Tau & 54 & 10.3 & 10.2 & 2.5 & 56 & 40.2 & 0.3\\
  GK Tau & 57 & 4.65 & 18.7 & 2.37 & 46 & 34 & 0.2\\
  DL Tau & 58 & 9.4 & 16 & 2.92 & 78 & 110.2 & 1.9 \\
  AA Tau & 63 & 8.2 & 11.3  & 1.94 & 69 & 14.2 & 0.3\\
  DN Tau & 65 & 6.0 & 8.1 & 1.93 & 30 & 15.7 & 0.1\\
  RW Aur & 80 & 5.3 & 17.2 & 3.00 & 37 & 45.9 & 2.0\\
  \hline

\end{tabular}}
\end{center}
\end{table}

\begin{figure}
\begin{center}
\resizebox{\hsize}{!}{\includegraphics[angle=0]{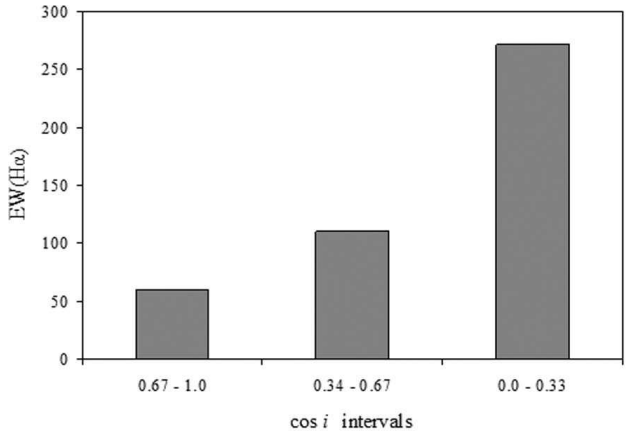}}
\caption{Histogram showing the CTTS H$\alpha$ equivalent width at
zero veiling averaged over the stars included in 3 bins of equal cos
$i$ intervals spanning the full range of view angles. A similar
result is found when distributing the stars over 5 bins.}
\label{histogram}
\end{center}
\end{figure}

In order to meaningfully compare the H$\alpha$ equivalent widths, we
computed their value at zero veiling using the procedure given in
\citet{2000AJ....119.1881A}. By doing this, we presumably eliminated
the effect of varying mass accretion rates in different stars on
their line emission strengths. We then constructed a histogram of
the average corrected EW(H$\alpha$) in equal cos $i$ intervals.
Fig.~\ref{histogram} shows the result for a 3-bin histogram, with
bins containing 6, 5, and 4 stars, respectively. A strong
correlation of the average corrected EW(H$\alpha$) with view angle
is apparent, readily understood in the framework of the
interpretation discussed above; as the view angle increases, the jet
contribution to the H$\alpha$ flux becomes more and more apparent as
a separate emission region, thus driving the equivalent width value
up. Note that the correlation is not due to the particular
distribution used in Fig.~\ref{histogram}, as we found that the
correlation remains strong even with 5 bins.

While a detailed investigation of inclination effects on CTTS
spectral lines is beyond the scope of this work on edge-on stars and
will be the topic of a forthcoming paper, Fig.~\ref{histogram}
clearly demonstrates that the view angle of CTTSs is one of the
parameters determining their emission properties. The two parameters
now known to play a role in CTTS line emission are the mass
accretion rate and the view angle of the system. The dispersion of
EW(H$\alpha$) within the single bins of Fig.~\ref{histogram}
indicates that (at least) a third yet unknown parameter must also
play a role in the complex CTTS line emission process. To conclude
this section, we note that since the equivalent widths of H$\alpha$
and forbidden lines vary with inclination, physical properties that
are usually derived from these quantities, such as the wind
mass-loss rate, must be viewed with some caution.

\subsection{Other permitted lines: Na\,{\sc i} D emission}
The interpretation of other permitted lines observed from the
spatially extended scattering region above and below the disks is
more difficult. Using the HST images and the known distance of the
Taurus association \citep[139 pc,][]{1999A&A...352..574B}, we find
that most of the light reaching us is scattered towards us at
heights of about 15 AU (HK Tau B), 35 AU (HV Tau C) and 55 AU (HH
30$^{*}$).  In all three cases most of the scattered light is
observed within a projected angle of $\pm 45^{\circ}$ of the pole
direction.  To connect these observational facts to the actual
scattering geometry requires some knowledge of disk geometry,
distribution of the scattering matter, and of the scattering
asymmetry parameter of the dust grains involved.

\subsubsection{Scattering from an extended envelope?}
 The spectroscopic results described above obviously
contain information on the scattering geometry. One of our main
findings has been the exceptionally narrow Na\,{\sc i} resonance
emission lines compared, e.g., to HL~Tau. As pointed out in Section
3.4, these lines were not detected in the parts of the jets observed
directly above and below the disks, which means that the outer
regions of the jets do not significantly contribute to these lines.
Moreover, the angular extent of the area from which these lines are
observed seems to correspond closely to that of the stellar
continuum. Thus the Na\,{\sc i} emission apparently originates in
the central part of the edge-on systems in the vicinity of the
central stars. Since significant emission in neutral resonance lines
is expected from volumes of cool intermediate-density gas, we expect
the innermost regions of the system to be the main source of
Na\,{\sc i} resonance line emission.

In the co-rotating magnetospheric accretion models generally assumed
today, the inner disk, the accretion flow, and the base of the jet
rotate approximately with the Keplerian  circular velocity of the
inner edge of the disk, so that we expect line emission from these
regions to show rotationally broadened profiles. Narrow profiles are
expected only if these regions are viewed perpendicular to the
rotational velocities, i.e., essentially pole-on. Obviously this
would be the case if the matter above and below the disk, which
scatters the radiation into our direction, is located close to the
disk symmetry axis. Such a geometry would also be consistent with
the narrowness of photospheric lines observed.

Scattering by matter vertically above and below the disk center is
to be expected if isotropically scattering dust grains are involved
and if (as predicted by protostellar model computations) the disks
are surrounded by dusty, but optically thin remnant protostellar
envelopes, or if the disk's matter density decreases slowly in the
vertical direction. If the scattering material is of protostellar
origin and is still in free fall, this would result in a small
blueshift of the scattered light. However at the vertical distances
considered here the corresponding velocities would be no more than a
few km s$^{-1}$ and would thus not be detectable in our
spectrograms.

The suggestion that edge-on disks could be effectively seen close to
pole-on not only appears counterintuitive but also apparently
contradicts current scattering light models. In the last few years,
edge-on T~Tauri disks have been modeled assuming $\alpha $-disks
irradiated by their central stars
\citep{1998ApJ...500..411D,1999ApJ...527..893D,2001ApJ...553..321D}
or using models with assumed parameterized radial and vertical
density distributions \citep[see
e.g.][]{1998ApJ...497..404W,2004ApJ...602..860W}. Theory, as well as
observations, suggests that grain growth does occur in the T~Tauri
disks \citep[see e.g.][]{2001ApJ...553..321D}. Disk matter could
therefore show much stronger forward scattering properties than
normal interstellar dust. In models assuming such strongly forward
scattering grains, the volume scattering the radiation from the disk
center into our direction will be shifted towards us, and thus we
will view the disk center at an intermediate angle, or under extreme
conditions, more or less edge-on. Such models are able to produce
intensity maps that agree reasonably well with observed images and
spectral energy distributions of edge-on systems.

However, as discussed by \citet{2004ApJ...602..860W} the many free
parameters of such models are not well constrained by comparisons
with the observed energy distributions and continuum light
distributions. Observations can be fitted reasonably well with a
wide range of dust properties and geometric parameters. Furthermore,
these models were developed chiefly for modelling the disk
near-infrared continuum emission and often do not take into account
the protostellar envelope and outflows that are associated with
these disks. \citet{2003ApJ...589..410S} indeed had to assume the
existence of a spherical envelope, in addition to the disk, to model
the I band image of HV~Tau~C. Such an additional envelope is not
needed to reproduce the K band image, as it maps a lower temperature
region. The possibility that dust contained in the envelope and jet
regions affects the scattering process in such a way that edge-on
objects are effectively seen from regions close to their rotation
axis, as far as their optical light is concerned, may thus not be as
implausible as seems at first. Further modelling of scattering in
YSO systems taking into account all their components (and not only
the disk) would be helpful to test this possibility quantitatively.

Are there other possibilities to understand how the narrow permitted
emission lines are formed in edge-on systems? In order to
investigate this question, we focus on the Na\,{\sc i} resonance
lines. In the following paragraphs, we briefly compare the Na\,{\sc
i}\,D\ lines observed in edge-on systems with those seen in other
CTTSs and discuss the various possible formation regions for these
resonance lines.

\subsubsection{Comparison with other CTTSs}

\begin{figure}
\begin{center}
\resizebox{\hsize}{!}{\includegraphics[angle=0]{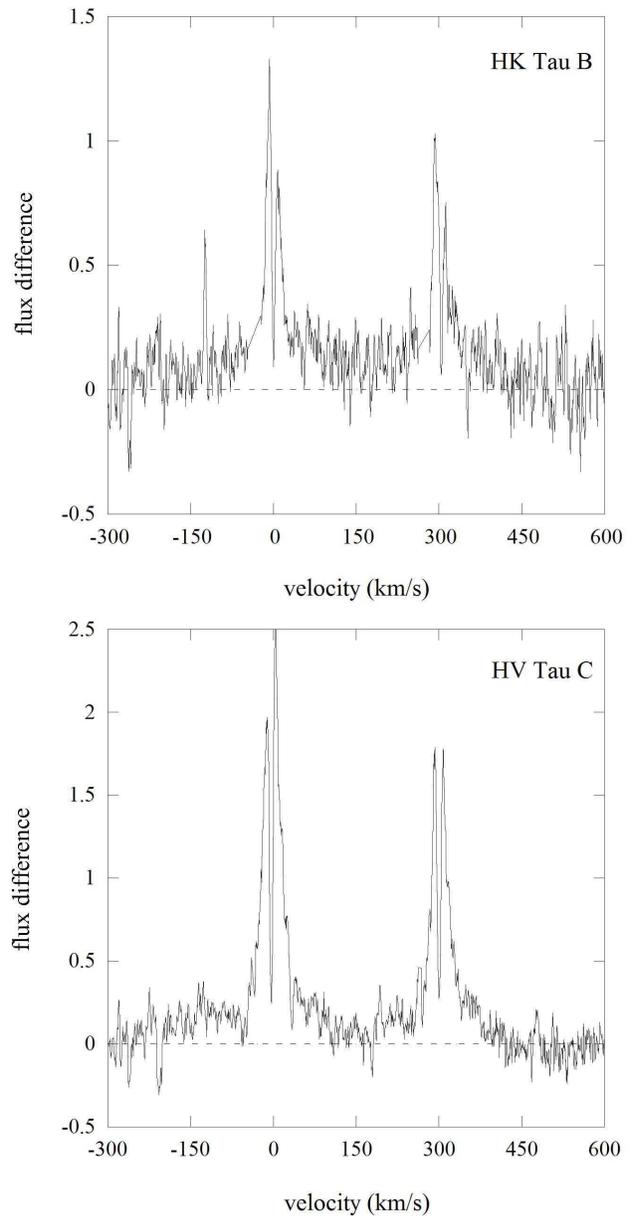}}
\caption{Residual Na\,{\sc i}\,D lines of HV~Tau~C and HK~Tau~B
after subtracting template photospheric absorption lines.}
\label{NaDResiduals}
\end{center}
\end{figure}

The following discussion is based on CTTS high resolution sodium
line profiles published by \citet{1994AJ....108.1056E},
 \citet{1998AJ....116..455M}, \citet{2000AJ....119.1881A}, and
 \citet{2002ApJ...571..378A}, so the reader is referred to these
 papers for a comparison with our data. Since some of the
 above authors do not present the observed profiles
 but residuals obtained by subtracting a template photospheric
 Na\,{\sc i}\,D\ from the observations, we provide the residuals
 for program stars HV~Tau~C and HK~Tau~C in Fig.~\ref{NaDResiduals}
 after subtracting from the spectra the photospheric lines of the template star LDN~1551-9. The
 extreme narrowness of the emission peaks
 is readily apparent from Figs.~\ref{profile_nad} and
 \ref{NaDResiduals}, while the residuals also display a wide, low flux
 level pedestal that extends up to about 100 km/s. An emission deficit in the blue wing of the
 line is also apparent in HV~Tau~C. The flux ratio of the blue to red line wings is 89\%.
 Our HH30* spectra are too noisy to allow for a meaningful integration of the flux,
 and the geocoronal sodium emission disturbs
 the weak emission line of HK~Tau~B, thus making it difficult to say
 whether a similar effect is present.

 A comparison with published Na\,{\sc i}\,D\ reveals both the wealth of
 profile shapes displayed by these resonance lines in the CTTS class and
 the almost unique position of edge-on stars in this class.
 Wide emission peaks are common (e.g.,
 DF~Tau, DG~Tau, DL~Tau), as are prominent
 blue-displaced absorption components (e.g., DF~Tau, DG~Tau, DO Tau)
 and shallow red-displaced absorption components (e.g., AA~Tau, DK Tau, DR~Tau).
 The only star in the sample of 27 CTTSs that displays similarly narrow emission peaks is the well-known
 star TW~Hya, the nearby CTTS that is seen almost pole-on
 \citep{2002ApJ...571..378A}.

 The question thus arises whether the narrowness of the Na\,{\sc i}\,D\ lines
 is a reliable indication that emission from the edge-on stars is
 effectively seen pole-on (as in TW~Hya)
 or whether the sodium line-forming region in our stars
 has properties that would result in an intrinsically
 narrow line width even at large view angle. To answer this question, we examine in turn
 the various possible regions for sodium line formation. Three
 possible regions come to mind: the magnetospheric accretion region, the
 innermost disk layers, and the inner disk wind region. In
 theory, these three regions can contribute to line emission
 from CTTSs, but in recent years attention has focused mainly on the
 magnetosphere as the main origin of CTTS line emission.

\subsubsection{Na\,{\sc i}\,D\ formation in the magnetospheric region}

The current paradigm for CTTSs assumes that permitted emission lines
are formed in the magnetospheric region; and detailed models were
computed by \citet{2001ApJ...550..944M}, who present a grid of
Na\,{\sc i}\,D\ residual profiles computed for various mass
accretion rates and magnetosphere properties for a view angle of
60\deg. These profiles often show an emission peak becoming wider at
low flux, and sometimes display shallow red-displaced absorption
components. In some cases, the emission peak becomes extremely
narrow as observed in our program stars. Interestingly, the Na\,{\sc
i}\,D\ lines of the near pole-on star TW~Hya were fitted reasonably
well by \citet{2002ApJ...571..378A} using one of these
magnetospheric profiles computed for a 60\deg view angle. One might
therefore conclude that the narrowness of sodium emission lines in
edge-on stars can readily be understood in the framework of the
current magnetospheric accretion paradigm.

One observed fact, however, is not reproduced in the framework of
this model. As noted above, the flux in the blue wing of the
Na\,{\sc i}\,D\ residual lines is smaller (by about 10\%) than in
the red wing of the HV~Tau~C lines, which in our sample has the best
signal-to-noise ratio. The signal-to-noise of our HK~Tau~B and HH30*
spectrograms does not allow for a meaningful integration of the
sodium line wings, but visual inspection of Fig.~\ref{profile_nad}
suggests that the same effect is present in the other edge-on stars.
This seems to indicate that blue-displaced material absorbs some of
the sodium line photons. When carefully studying theoretical
emission profiles formed in magnetospheric accretion regions,
however, one sees that the situation is \emph{always} reversed;
i.e., the red-ward line flux is always smaller than the blue-ward
flux, reflecting absorption in the infalling magnetospheric region.
We thus conclude that the outflow is likely to play a role in the
formation of Na\,{\sc i}\,D\, at least in HV~Tau~C and possibly in
other edge-on stars.

\subsubsection{Na\,{\sc i}\,D\ formation in the innermost disk region}

\begin{figure}[tbp]
\begin{center}
\resizebox{\hsize}{!}{\includegraphics[angle=0]{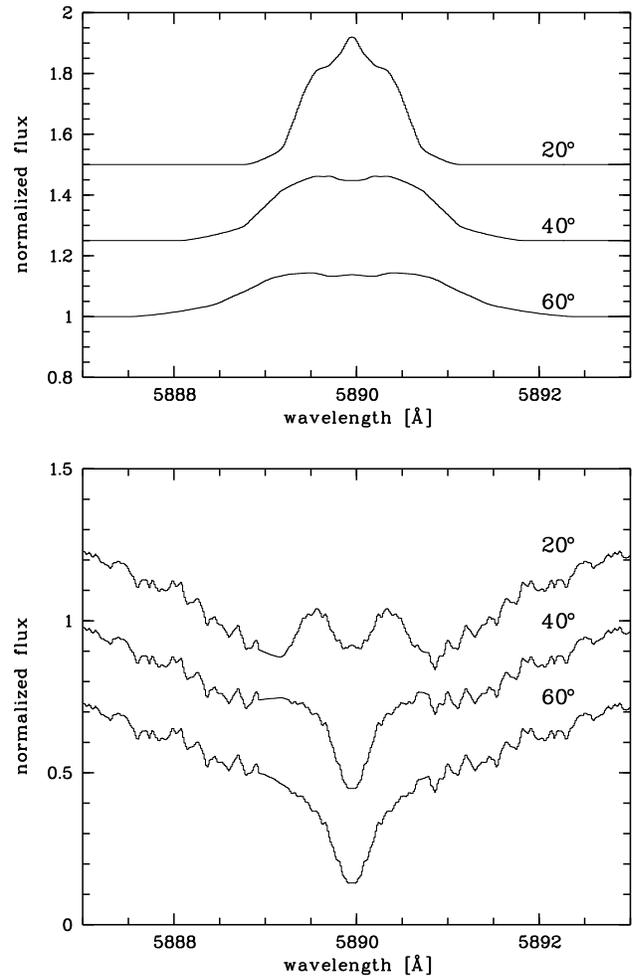}}
\end{center}
\caption{Na\,\textsc{i}\,D1 line profiles formed in an accretion
disk seen at different view angles ($20^{\circ}$, $40^{\circ}$, and
$60^{\circ}$). Panel (a): Line residuals as defined in the text.
Panel (b): the photospheric Na\,\textsc{i}\,D1 absorption line is
taken into account in the computation.} \label{DiffAngles}
\end{figure}

The innermost regions of irradiated accretion disks reach
temperatures in the $2000 - 3000$ K range, where sodium emission can
occur if the gas density is sufficient. We use a 2D line transfer
code developed for investigating ejection/accretion structures
(Bertout et al., in preparation) to compute line formation in the
inner parts of an $\alpha$ accretion disk with reprocessing taken
into account as in \citet{2000A&A...363..984B}, and found that
narrow, doubly-peaked emission lines could be formed in this region,
provided the disk is seen close to pole-on. As soon as the view
angle becomes larger, the emission lines become wider as expected on
the basis of the quasi-Keplerian velocity field present in the disk.
Fig.~\ref{DiffAngles} presents line profiles originating from the
innermost disk region as seen by observers looking at the system
from different view angles. The disk parameters are the same as
given in Table~\ref{TablePrms} except for the mass accretion rate,
which is larger by a factor of ten. View angles are respectively 20,
40, and 60 degrees. Panel (a) of Fig.~\ref{DiffAngles} shows the
residual line emission as defined above, and Panel (b) shows the
disk emission when the Na\,\textsc{i}\,D1 photospheric background
flux is taken into account. Clearly, the extended wings due to disk
emission which are seen in the Na\,\textsc{i}\,D1 line even at
moderate view angles, are not compatible with observations.

In the framework of the (admittedly simple) disk model investigated
here, we found that the Na\,{\sc i}\,D line emission from the disk
alone is insufficient to explain the observed flux unless the
mass-accretion rate is several $10^{-7}$ M$_{\odot}$yr$^{-1}$, which
is not compatible with the moderate veiling observed in edge-on
systems. A potential way out of this difficulty is to consider the
strong anisotropy of stellar light reprocessing, which would result
in a temperature inversion at the disk surface \citep[see, e.g.,
][]{1991ApJ...383..814M}. It is obvious that such a "chromospheric"
disk layer will somewhat increase emission in the optical resonance
lines. Thus, more detailed disk models that take surface heating
into account in a more accurate way need to be investigated in order
to precisely calibrate the line emission expected from accretion
disks. It appears, nevertheless, unlikely that strong resonance
lines can be entirely formed in the disk at the moderate accretion
rates displayed by many CTTSs.

We thus rule out disk emission alone as a possible explanation of
the observed narrow Na\,{\sc i}\,D\ lines. We note, however, that
disk emission contributes to filling up the deep photospheric
absorption features (Fig.~\ref{DiffAngles}), even at moderate disk
accretion rates, and  is thus a needed component for explaining the
emission lines formed in the disk wind models discussed in the next
section.

\subsubsection{Na\,{\sc i}\,D\ formation in the inner disk wind}

In the disk wind acceleration region located close to the disk
surface, the poloidal velocities are expected to be small, while the
toroidal velocities are comparable to the Keplerian velocity of the
disk region where the wind is launched \citep [see, e.g., ]
[]{2004A&A...416L...9P}. One would thus tend to expect the lines
formed there to be wide as soon as the view angle is moderate,
similar to the disk case. There is, however, an additional factor to
take into account, the acceleration and collimation of the outflow
expected to occur within a few AUs. If collimation of the outflow
takes place on small scales, it could well be that the range of
observed radial velocities could be narrowed even at high
inclination angles.

In order to test this possibility, we used our 2D line formation
code to compute line profiles emerging from a disk/jet toy model
that is set up as follows. The disk is modeled as in the previous
paragraph. We then assume that the jet radial velocity increases as
$v \propto r^\delta$ from the local sound speed at the inner disk
radius to a maximum velocity $V_\mathrm{jet}$ that is reached at the
outside boundary of the computational domain, a sphere of radius
1AU. For computational simplicity, the rotational velocity component
in the jet $V_\phi$ is assumed to be equal to a constant fraction of
the Keplerian velocity at the inner disk radius, and the base of the
jet is assumed to fill up the full solid angle not occupied by the
disk, except for the innermost cylinder between the rotation axis
and the inner disk radius, which we take to be hollow. We use a
latitude-dependent density law to mimic collimation along the
rotation axis. More specifically, we assume that the jet density
varies as $\rho\left( r,\theta\right) \thicksim \rho\left(
r,0\right) \left[ 1-\left( \sin\theta/\sin\theta_\mathrm{D}\right)
^{\beta}\right] $ where $\theta$ is the angle between the jet axis
and the radius vector $\overrightarrow{r}$ and where
$\theta_\mathrm{D}$ is the disk's opening angle. The radial
variation of density follows from the condition of mass
conservation, assuming that the mass-loss rate in the jet is equal
to a fraction $\gamma$ of the disk mass-accretion rate. The jet is
assumed to be isothermal with temperature $T_\mathrm{jet}$. Other
parameters of the system are the stellar radius $R_\mathrm{star}$,
the stellar mass $M_\mathrm{star}$, its effective temperature
$T_\mathrm{eff}$, the disk inner radius $R_\mathrm{D}^{\min}$ and
outer radius $R_\mathrm{D}^{\max}$, its mass-accretion rate
$\dot{M}_\mathrm{D}$, and the viscosity parameter $\alpha$.

\begin{figure}
\begin{center}
\resizebox{\hsize}{!}{\includegraphics[angle=0]{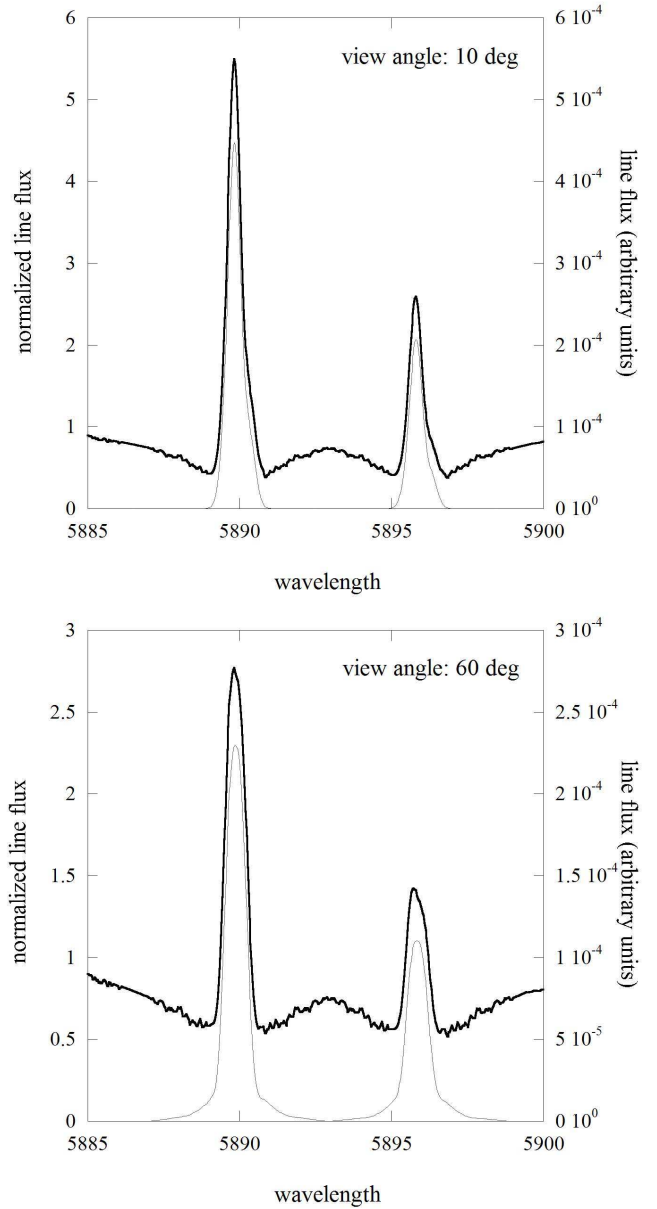}}
\caption{Na\,{\sc i} D line profiles formed in the toy model
described in the text. The darker solid line shows the line flux
normalized to the continuum in the vicinity of the lines, while the
lighter solid line shows the residual line flux once the
photospheric contribution is subtracted. Parameters for this
computations are those of Model 1 in Table~\ref{TablePrms}. In
particular, a modest collimation of the wind is assumed here.}
\label{model1}
\end{center}
\end{figure}

\begin{figure}
\begin{center}
\resizebox{\hsize}{!}{\includegraphics[angle=0]{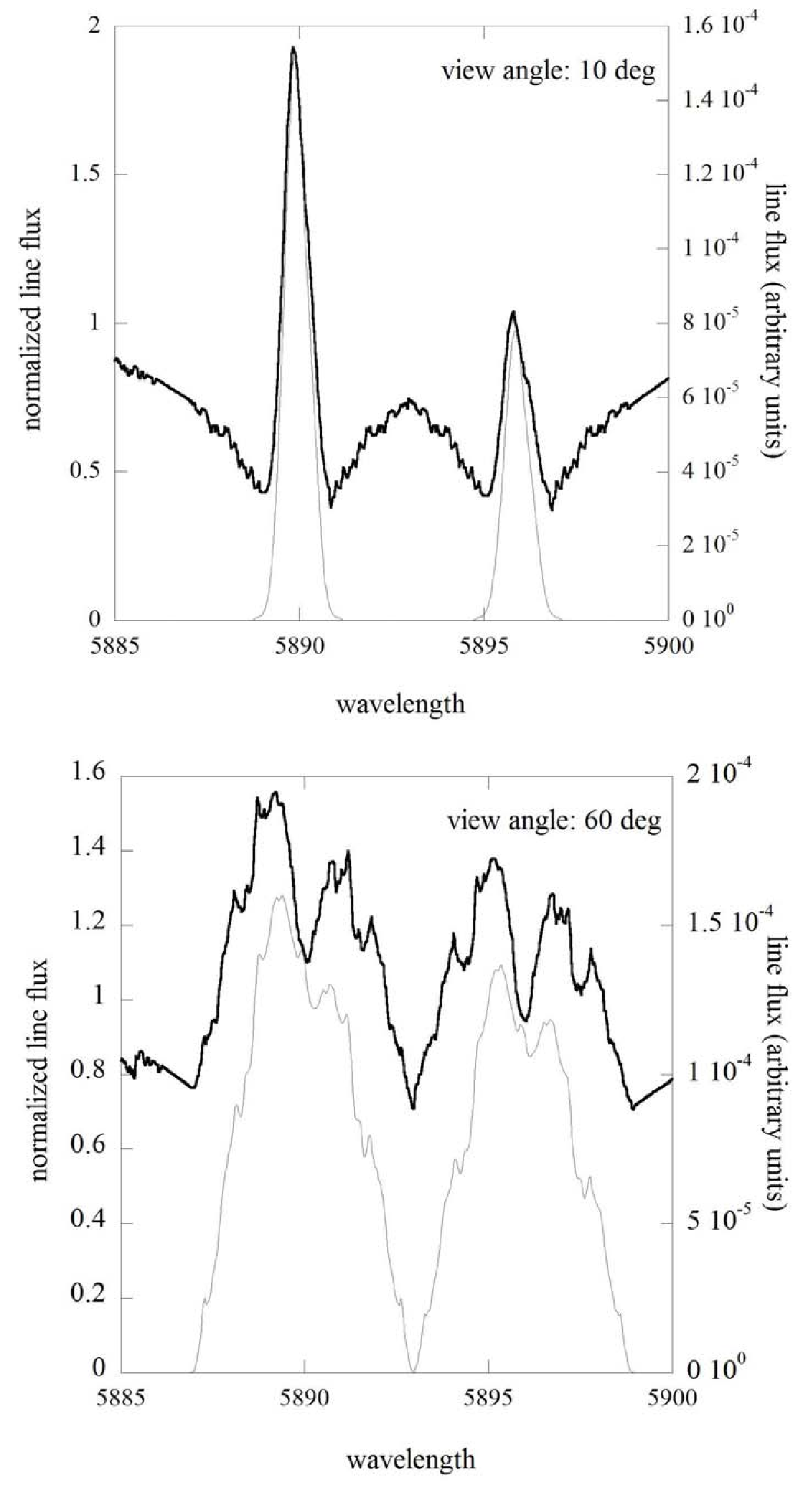}}
\caption{Same as Fig.~\ref{model1} but for the parameters of Model~2
in Table~\ref{TablePrms}. Here, the disk wind is uncollimated.}
\label{model2}
\end{center}
\end{figure}

Using this toy model, we found that line profiles resembling
observed ones were easily produced for system parameters that are
typical of CTTSs, and we computed two sets of models, shown in
Figs.~\ref{model1} and~\ref{model2}. The first one is for modest jet
collimation ($\beta=2$) and the second one is uncollimated
($\beta=0$). Table~\ref{TablePrms} presents the parameters relevant
to the computed sets of profiles. With the modest disk-accretion
rate assumed here, the main contribution to Na\,{\sc i} line
emission comes from the base of the wind. Thus, in accordance with
the observations, the fast-moving outer part of the jet does not
contribute to the Na\,{\sc i}\,D lines.

\begin{table}[tbp] \centering
\renewcommand\arraystretch{1.1}
\begin{tabular}
[c]{ccc}\hline\hline & Model 1 & Model 2\\
\hline
$R_\mathrm{star}$ & \multicolumn{2}{c}{$2$ R$_{\odot}$}\\
$M_\mathrm{star}$ & \multicolumn{2}{c}{$0.5$ M$_{\odot}$} \\
$T_\mathrm{eff}$ & \multicolumn{2}{c}{$3500$ K}\\
$R_\mathrm{D}^{\min}$ & \multicolumn{2}{c}{$2$ $R_\mathrm{star}$} \\
$R_\mathrm{D}^{\max}$ & \multicolumn{2}{c}{$1$ AU} \\
$\dot{M}_\mathrm{D}$ & \multicolumn{2}{c}{$4\cdot10^{-8}$
M$_{\odot}$yr$^{-1}$}\\
$\alpha$ & \multicolumn{2}{c}{$10^{-2}$}\\
$\beta$ & $2$ & $0$\\
$\gamma$ & $5\cdot10^{-2}$ & $3\cdot10^{-2}$ \\
$T_\mathrm{jet\text{ }}$ & \multicolumn{2}{c}{$10^{4}$ K} \\
$V_\mathrm{jet}$ & \multicolumn{2}{c}{$100$ km s$^{-1}$} \\
$V_\phi$ & \multicolumn{2}{c}{$15$ km s$^{-1}$ }
\\\hline
\end{tabular}
\caption{Computation parameters for the lines shown in
Figs.~\ref{model1} and~\ref{model2}. The various parameters are
defined in the text.} \label{TablePrms}
\end{table}

We found that the Na\,{\sc i} profiles similar to those observed can
be explained by a model combining a disk accretion rate in the range
$1-5\cdot10^{-8}$ M$_{\odot}$yr$^{-1}$, which are typical values for
moderate CTTSs, with a wind expelling a few \% of the accreted mass.

Collimation of the wind is found to be of crucial importance in
determining the shape of profiles formed in our toy model. Emission
profiles formed in a collimated wind (Model 1) remain narrow at all
view angles, even for the moderate amount of collimation that we
used in our toy model. This occurs because the emitting gas is
concentrated along the rotation axis so that the range of projected
velocities seen by the observer as originating from the emitting
volume is narrow. As explained in the previous paragraph, emission
from the inner disk layers provides the wide basis seen in the
residual profile at 60\deg view angle in Fig.~\ref{model1}, which
fills up partly the photospheric absorption.

In an uncollimated wind (Model 2), by contrast, the projected
outflow velocities in the observer's direction cover a wider range
at all view angles and, together with the underlying disk emission,
contribute to the widening of the overall profile. We note the
resemblance between the sodium lines observed in HL~Tau and the
model profiles for 60\deg. In this uncollimated wind, narrow
profiles can be achieved only when disk emission is suppressed,
i.e., when the disk is seen close to pole-on.

To the question we asked ourselves at the beginning of this section
we can now answer that there are indeed some disk wind geometries
that produce narrow Na\,{\sc i}\,D emission lines for all view
angles. In fact, these profiles bear strong resemblances to profiles
formed in magnetospheric accretion regions, as the residual profiles
shown in Fig.~\ref{model1} indicate. Both models predict narrow
residual profiles with an additional wide basis when seen at 60\deg.
In addition, the wind residual profiles formed in our toy model
exhibit a modest flux deficit on the blue side of the lines that is
presumably due to self absorption. More specifically, the flux in
the blue side of the lines seen is about 80\% of the flux in the red
line side. As noted earlier, a similar effect is observed in the
sodium line residuals of HV~Tau~C.

There is nevertheless a problem with the idea that the Na\,{\sc
i}\,D lines might be formed in such a wind model. Since the view
angle hardly affects the profile shape of the computed lines, one
should observe many stars with such narrow profiles, if this
physical mechanism is the dominant one in sodium line formation. As
noted above, this is not the case; among the many CTTSs observed at
high spectral resolution, only the edge-on stars and TW~Hya were
found to exhibit narrow Na\,{\sc i}\,D emission peaks. In many
cases, the observed profiles are wide, as seen e.g. for HL~Tau in
Fig.~\ref{profile_nad}. The same criticism probably applies to the
published sodium profiles formed in magnetospheric models, which do
not seem able to produce wide emission components even at high
inclination.

It could well be, however, that more realistic disk wind models will
be able to reproduce different resonance line shapes for the same
view angles depending, for example, on the wind collimation degree
or the accretion/ejection geometry. In this context, the current
lack of physically consistent models for the accretion/ejection
region and jet is a real problem for interpreting YSO emission line
profiles. The development of realistic models for accretion/ejection
structures is under way \citep[e.g.,][]{2004A&A...416L...9P} and
appears to be the main key to further understanding YSO physics. The
results reported here are nevertheless encouraging because they
demonstrate that a simple toy model can easily account for the bulk
properties of the Na\,{\sc i} resonance line emission with a set of
parameters relevant to CTTS systems.

\section{Conclusions}

As pointed out above, our observations have shown that the observed
edge-on disks in the Taurus star formation region are actually CTTSs
where the line-of-sight to us by chance coincides with the plane of
the proto-stellar (and pre-planetary) accretion disk. Thus, our
observations provide further support to the current theoretical
concepts of low-mass star formation which assume that the CTTSs are
pre-main sequence stars surrounded by both cool, dusty accretion
disks and remnants of their proto-stellar envelopes. The different
amounts of veiling observed and the different jet strengths of the
three observed edge-on disks indicate different mass accretion rates
of the central T~Tauri stars, or possibly different evolutionary
stages. But otherwise these stars show no detectable systematic
differences from other known CTTSs apart from the different viewing
angle. We conclude that all classical T~Tauri stars very likely show
the morphology of our targets when viewed edge-on. That edge-on
T~Tauri stars have met with little attention so far is obviously
explained by the fact that at the same distance they appear much
fainter than their less inclined counterparts. With the new large
telescopes and better detectors probably many more of these objects
will be found in galactic star formation regions.

For a comparison with theoretical models, edge-on T~Tauri stars have
the significant advantage that the inclination of the disk and jet
systems is known. Because of the high spectral resolution of the
{\sc Uves} spectra and the narrow profiles of the forbidden lines in
our edge-on objects, the H$\alpha$, He\,{\sc i}, and forbidden lines
show interesting details that could not be detected in earlier
spectra. As described in Section 3, our spectra show two different
components of line emission, which originate from different regions
and under different physical conditions in the T~Tauri systems. They
also demonstrate beyond any doubt that jet emission contributes not
only to forbidden emission but also to permitted lines such as
H$\alpha$ and He\,{\sc i}, which suggests that the view angle plays
a role in the appearance of CTTS emission lines. In Section 4.2, a
comparison of H$\alpha$ equivalent widths in a small sample of
Taurus-Auriga CTTSs with known view angles showed that the H$\alpha$
equivalent widths increase on average with the system's view angle
after correcting for the effect of veiling

As discussed above, we conclude that there are two possible ways to
interpret the optical light reaching us from the center of these
systems: (a) it could be scattered in our direction by matter
located along the rotation axis, or (b) it could originate mainly at
the base of a disk wind showing some degree of collimation, in which
case light scattering occurs in the disk atmosphere as usually
assumed. In Section 4 we have demonstrated that with this second
hypothesis the modeling of the resonance line emission from the disk
and the base of the jet can reproduce the observed narrow Na\,{\sc
i} emission lines of our edge-on T~Tauri stars. However, further
modeling of scattering in YSO envelopes and of line formation in
realistic disk wind models is needed to make a definitive choice
between these conclusions.

Comparing our spectra of the edge-on stars with that of the T~Tauri
star HL~Tau we find similarities as well as marked
differences. However, all observed differences are readily explained
by the edge-on aspect of our program objects. In view of this
result we regard our study as a promising first step towards a
thorough investigation of orientation effects in T~Tauri spectra.

\acknowledgements{ It is a pleasure to thank the ESO Paranal
Observatory staff for carrying out for us the service mode
observations on which this paper is based. Many thanks also to
Sylvie Cabrit and referee Christopher Johns-Krull for carefully
reading our manuscript and for making many valuable comments. We are
indebted to J\'er\^ome Bouvier for useful comments and for
communicating to us his compilation of rotation periods and radial
velocities for T Tauri stars. I.A. gratefully acknowledges the award
by the French Ministry of Education of a Prix Gay Lussac-Humboldt,
which resulted in a very enjoyable stay at the IAP, Paris, in the
course of which this work was initiated. I.A. also wishes to thank
C.B., and the director and staff of the IAP for their kind
hospitality. This joint research project was supported in part by
the European Research Training Network ``The Origin of Planetary
Systems'' (PLANETS, Contract Number HPRN-CT-2002-00308). }

\bibliographystyle{aa}
\bibliography{2217}

\end{document}